\documentclass[%
 reprint,
superscriptaddress,
nofootinbib,
 amsmath,amssymb,
 aps,
]{revtex4-2}
\usepackage{graphicx}
\usepackage{dcolumn}
\usepackage{bm}
\usepackage[version=4]{mhchem} 
\usepackage{subfig}
\usepackage{float}
\usepackage{caption}
\usepackage{multirow}
\usepackage[colorlinks,
            linkcolor=red,
            anchorcolor=blue,
            citecolor=blue
           ]{hyperref}

\graphicspath{ {figure/} }
\begin{document}

\title{Precise measurement of the cosmic-ray spectrum and \texorpdfstring{$\left \langle \ln A \right \rangle$}{mean mass} by LHAASO  --- connecting the Galactic to the extragalactic components}

\author{Xing-Jian Lv}
 \affiliation{%
 Key Laboratory of Particle Astrophysics, Institute of High Energy Physics, Chinese Academy of Sciences, Beijing 100049, China}
\affiliation{
 University of Chinese Academy of Sciences, Beijing 100049, China 
}%
 \author{Xiao-Jun Bi}
\affiliation{%
 Key Laboratory of Particle Astrophysics, Institute of High Energy Physics, Chinese Academy of Sciences, Beijing 100049, China}
\affiliation{
 University of Chinese Academy of Sciences, Beijing 100049, China 
}%
\affiliation{Tianfu Cosmic Ray Research Center, 610000 Chengdu, Sichuan,  China}

\author{Kun Fang}
\affiliation{%
 Key Laboratory of Particle Astrophysics, Institute of High Energy Physics, Chinese Academy of Sciences, Beijing 100049, China}
 
\author{Yi-Qing Guo}
\affiliation{%
 Key Laboratory of Particle Astrophysics, Institute of High Energy Physics, Chinese Academy of Sciences, Beijing 100049, China}
 
 \author{Hui-Hai He}
\affiliation{%
 Key Laboratory of Particle Astrophysics, Institute of High Energy Physics, Chinese Academy of Sciences, Beijing 100049, China}
\affiliation{
 University of Chinese Academy of Sciences, Beijing 100049, China 
}
\affiliation{Tianfu Cosmic Ray Research Center, 610000 Chengdu, Sichuan,  China}

\author{Ling-Ling Ma}
\affiliation{%
 Key Laboratory of Particle Astrophysics, Institute of High Energy Physics, Chinese Academy of Sciences, Beijing 100049, China}
\affiliation{Tianfu Cosmic Ray Research Center, 610000 Chengdu, Sichuan,  China}
 
 \author{Peng-Fei Yin}
\affiliation{%
 Key Laboratory of Particle Astrophysics, Institute of High Energy Physics, Chinese Academy of Sciences, Beijing 100049, China}
 
\author{Qiang Yuan}
\affiliation{Key Laboratory of Dark Matter and Space Astronomy \& Key Laboratory of Radio Astronomy, Purple Mountain Observatory, Chinese Academy of Sciences, 210023 Nanjing, Jiangsu, China}

\author{Meng-Jie Zhao}
 \affiliation{%
 Key Laboratory of Particle Astrophysics, Institute of High Energy Physics, Chinese Academy of Sciences, Beijing 100049, China}
\affiliation{
China Center of Advanced Science and Technology, Beijing 100190, China 
}%


\date{\today}

\begin{abstract}
Recently LHAASO Collaboration gives precise measurements of cosmic rays (CR) all particle energy spectrum and mean logarithmic mass $\left \langle \ln A \right \rangle$ from 0.3 PeV to 30 PeV. Combining the CR measurements by AMS-02 and DAMPE in space and that by LHAASO and Auger on the ground we construct a model to recover all these measurements from tens of GeV to tens of EeV. We find the LHAASO measurement is crucial in the model construction by connecting the Galactic component to the extragalactic component. The precise measurements of CR spectra for individual species by AMS-02 and DAMPE together with the newest LHAASO results clearly indicates three Galactic CR components, that is, a soft low energy background, a hard high energy component, and a local source contribution. However, the LHAASO data show that above $\sim 10^{16}$ eV a nonnegligible extragalactic component must be included. Combining the Auger results and the LHAASO results we figure out the extragalactic CRs which need at least two components at lower and higher energies. Thanks to the precise measurements by LHAASO the constraints on the model parameters are quite stringent.
The spectra features and mass measurements in all energy range are all well reproduced in the model. 

\end{abstract}
\maketitle


\section{\label{sec:level1}INTRODUCTION}
More than a century has passed since Victor Hess's groundbreaking balloon experiment in 1912~\cite{Hess:1912srp}, yet the origins and characteristics of cosmic rays (CRs) remain enigmatic~\cite{BeckerTjus:2020xzg}. Cosmic rays span an immense energy range, from below $10^8$ eV up to more than $10^{20}$ eV. By our current understanding, most of the CRs up to approximately $10^{17}$ eV originate within our Galaxy, likely linked to supernova events, while CRs with energies higher than about $10^{18}$ eV are generally considered to be of extragalactic origin, produced from more extreme processes like active galactic nuclei or gamma-ray bursts. The precise energy threshold marking the transition from galactic to extragalactic sources remains a subject of ongoing debate~\cite{AlvesBatista:2019tlv}.

Based on the theory of diffusive shock acceleration~\cite{LOC:Drury_1983} and CR propagation as a diffusive process~\cite{Strong:2007nh}, cosmic rays are traditionally understood to exhibit a power-law spectrum. The feature known as `the knee' is ascribed to Peter's cycle~\cite{Peters:1961mxb}, reflecting the maximum energy attainable by Galactic CR sources. However, recent space-born experiments have unveiled several unexpected features in the CR spectrum. These anomalies include a hardening at $\sim200$ GV across most of the primary CR spectra~\cite{Panov:2006kf, Panov:2009iih, Ahn:2010gv, Yoon:2017qjx, PAMELA:2011mvy, AMS:2015azc, AMS:2015tnn}, a softening at $\sim10$ TV in the spectra of protons and helium~\cite{Yoon:2017qjx, Atkin_2018, DAMPE:2019gys, Alemanno:2021gpb}, and a subsequent hardening at $\sim100$ TeV in the combined proton and helium spectrum\cite{dampecollaboration2023measurement}. Concurrently, ground-based experiments have not only validated the well-established knee at roughly 4 PeV, the ankle around 5 EeV, and the suppression at 40 EeV, while also uncovering new phenomena: a low-energy ankle at \(2 \times 10^{16}\) eV~\cite{IceCube:2013ftu,Prosin:2016rqu} and a second-knee at around \(10^{17}\) eV (see~\cite{Bergman:2007kn} for a review).

Recently, the Large High Altitude Air Shower Observatory (LHAASO) published the measurement of the all-particle CR energy spectrum  and the mean logarithmic mass of CRs with unprecedented accuracy in 0.3-30 PeV~\cite{lhaaso2024allparticle}. The mean logarithmic mass shows an unexpected nonmonotonic change with energy. 


Phenomenological studies have been conducted utilizing data available at different times~\cite{Hoerandel:2002yg, Zatsepin:2006ci, Hillas:2006ms, Gaisser:2013bla, Stanev:2014mla, Thoudam:2016syr, Dembinski:2017zsh, Guo:2017tle, Yue:2019sxt}. 
Ref.~\cite{Dembinski:2017zsh} compiled the most up-to-date data available then, spanning a broad energy spectrum. However, their approach, a Global Spline Fit (GSF), lacks physical insights into the specific populations contributing to the observed spectrum and compositions. 
Consequently, this study revisits the phenomenological modeling of cosmic rays, covering a comprehensive energy range from tens of $\mathrm{GeV}/Z$, an energy range where solar modulation effects are negligible~\cite{Potgieter:2013pdj}, up to $10^{11}$ GeV, the highest detectable energies. 

Motivated by the recent LHAASO result we construct a comic ray model which benefits greatly from the precise LHAASO measurements on the spectrum and $\left \langle \ln A \right \rangle$ around the knee. The model leads us to a comprehensive insight into the origins of cosmic rays through the whole energy range.
At the lower energy limit of its measurements, the impact of the local source, responsible for the spectral bump detected by space-borne experiments around 10 TV \cite{DAMPE:2019gys, Alemanno:2021gpb}, remains pronounced. Notably, the cutoff in the iron component from the local source contributes to the observed decrease in $\left \langle \ln A \right \rangle$ by LHAASO for the first time. For the higher energies, the precise measurements of both the all-particle spectrum and $\left \langle \ln A \right \rangle$ around the knee strongly indicate a `light knee' resulting from the sequential cutoff of galactic protons and helium. At the uppermost energies of LHAASO, the emergence of the low-energy ankle helps  to constrain the low energy end of extragalactic contributions.


A detailed explanation of our models and the datasets employed for calibration is presented in Section~\ref{sec methology}. The results and their implications are discussed in Section~\ref{sec results}. Finally, we summarize our findings in Sec.~\ref{sec:conclusion}.

\section{The model construction}\label{sec methology} 

We first compile all the CR data from the space based experiments AMS-02, DAMPE and CALET to the ground based experiments LHAASO, HAWC and Auger. We assume the spectra discrepancy between any two measurements with overlap energies is attributed to the absolute energy calibration of the two experiments. By slightly adjusting the energy scale we calibrate the absolute energies of all experiments with AMS02. Then we get a CR spectrum from GeV to tens of EeV. Our model is then constructed to reproduce the whole energy spectrum.

It is shown clearly that in order to reproduce the spectrum three galactic CR components are necessary: a local source and two background populations, referred to as Population I (Pop. I) and Population II (Pop. II). Meanwhile, the extragalactic contributions are composed of two populations, i.e. the Low Energy (LE) and the High Energy (HE) populations. 
The spectrum of each population is assumed to be a power-law with exponential cutoff, as detailed in Sec.~\ref{sec: spectrum}. The cutoff energies of different species of each population depend on the atomic number $Z$, a natural consequence if the cutoff originates from acceleration or propagation processes~\cite{Hoerandel:2002yg}.


\subsection{Data Compiled\label{sec: data}}
For space-borne experiments, we opt to use the latest results by AMS-02~\cite{AMS:2021nhj}, DAMPE~\cite{DAMPE:2019gys} and CALET~\cite{CALET:2022vro}. The AMS-02 dataset is presented in terms of rigidity, whereas the DAMPE and CALET data are expressed as total kinetic energy. To compare with ground-based experiments, a conversion to total energy is needed. We assume that the flux is dominated by a single isotope except for helium\footnote{This assumption disregards isotopic variations because, for proton, $\ce{^{1}H}$ dominates over $\ce{^{2}H}$ by orders of magnitude\cite{PAMELA:2015kyy}, and for elements heavier than helium, such as carbon, the mass differential between isotopes (e.g., $\ce{^{12}C}$ vs. $\ce{^{13}C}$) is relatively minor. Helium is an exception, as its isotopes, $\ce{^{3}He}$ and $\ce{^{4}He}$, not only share comparable fluxes but also exhibit a substantial relative mass difference.}, where the fitted function of AMS-02 \cite{AMS:2019nij} for the $\ce{^{3}He}/\ce{^{4}He}$ ratio is used to achieve an accurate conversion. Furthermore, the DAMPE measurement of proton plus helium (p+He) is converted using the ratio derived from the DAMPE proton and helium spectrum. We checked that adopting a 50/50 split, as done in the DAMPE p+He study, does not significantly alter the results. Additionally, when measurements from AMS-02, DAMPE, or CALET are unavailable, data from NUCLEON~\cite{GREBENYUK20192546} are employed. 

Space-borne experiments lack sufficient statistics above $10^5$ GeV, necessitating the use of ground-based experiments due to their larger acceptance. Given LHAASO's pivotal role in connecting the galactic and extragalactic cosmic rays , coupled with its unprecedented accuracy in this energy range, we start from LHAASO measurements of all-particle spectrum and $\langle\ln A\rangle$. We then select experiments whose datasets can be aligned with LHAASO's through energy rescaling. Following this approach, our data compilation includes the all-particle spectrum and mean logarithmic mass, $\langle\ln A\rangle$, measurements from TUNKA-133~\cite{PROSIN201494} and the Pierre Auger Observatory~\cite{PierreAuger:2021hun}, alongside the all-particle and p+He data from the High Altitude Water Cherenkov (HAWC)~\cite{HAWC:2021ubt}. After the construction of the model, we also briefly discuss the potential implications of using data that were not initially included, such as IceCube-IceTop~\cite{IceCube:2020yct},  KASCADE~\cite{Finger:2011bia}, KASCADE-Grande~\cite{Apel:2013uni}, Telescope Array (TA)~\cite{Ivanov:2020rqn}, the TA Low Energy Extension  (TALE)~\cite{TelescopeArray:2018bya}, and Tibet-III~\cite{TIBETIII:2008qon}
\footnote{We use the latest version of QGSJETII as provided by the experimental groups, except in the case of Auger data, for which we choose the EPOS-LHC results due to the Auger collaboration's demonstration that outcomes derived using QGSJETII were unphysical~\cite{PierreAuger:2023bfx}.}. 

Data taken by different experiments often have systematic discrepancies. In previous works~\cite{Hoerandel:2002yg, Gaisser:2013bla, Stanev:2014mla, Dembinski:2017zsh}, these discrepancies are attributed to energy calibration biases. In the work, we rescale a measured energy as $E_b\to\delta_b\times E_b$, aiming to minimize the deviation between experiments $a$ and $b$ which is quantified by
\begin{equation}
  \chi^2_{ab}(\delta_b)\equiv \sum_i 
  \frac{(\mathcal{D}_{a}(E_i)-\mathcal{D}_{b}((\delta_b E )_i))^2}{\sigma^2_{a}(E_i)+\sigma^2_{b}((\delta_b E )_i)} \;,
\end{equation}
where the summation runs over the overlapping energy range between experiment $a$ and $b$. Given that the energy bins are different for different experiments, we use the energy bins of experiment $a$ as a reference and accordingly interpolate both $\mathcal{D}_{b}(\delta_b E_b )$ and $\sigma^2_{b}(\delta_b E_b )$.
The AMS-02 data serve as the benchmark for energy calibration, benefiting from its precise energy calibration enabled by its permanent magnet. Subsequently, the CALET, DAMPE, and NUCLEON datasets are calibrated against the corresponding AMS-02 data. The minimization package iminuit\footnote{\url{https://zenodo.org/records/10638795}}~\cite{James:1975dr} is adopted to optimize $\delta$ for space-borne experiments, with results summarized in Table~\ref{tab:experiments}. 

To calibrate the energy scale of ground-based experiments, affected by large systematic uncertainties, we use the positions of spectral features, such as the knee and the second-knee, to obtain $\delta$. Firstly, we calibrate HAWC's energy scale using the proton-helium (p+He) spectrum as determined by DAMPE~\cite{dampecollaboration2023measurement} and HAWC~\cite{HAWC:2022zma}. Then, we compare the all-particle spectrum of LHAASO~\cite{lhaaso2024allparticle} against that of HAWC~\cite{HAWC:2021ubt}. Following this, TUNKA-133's all-particle spectrum~\cite{PROSIN201494} is contrasted with LHAASO's. Finally, the all-particle spectrum of Auger~\cite{PierreAuger:2021hun} is compared with TUNKA-133. 
Energy scales for experiments not directly involved in our model's construction are then adjusted through comparison of the all-particle spectrum.
The calibration parameters for each ground-based experiment are summarized in Table \ref{tab:ground}.

\begin{table}[]
\centering
\begin{tabular}{ccc}
\hline
Experiment        & $\delta$ & Reference\\
\hline
AMS-02            & 1.00 & \cite{AMS:2021nhj}\\
DAMPE proton      & 1.00 $\pm$ 0.01 &\cite{DAMPE:2019gys}\\
DAMPE helium      & 1.02 $\pm$ 0.02 &\cite{Alemanno:2021gpb}\\
DAMPE p+He        & 1.02 $\pm$ 0.02 &\cite{dampecollaboration2023measurement}\\
NUCLEON C\&O       & 1.01 $\pm$ 0.02 &\cite{GREBENYUK20192546}\\
NUCLEON Ne\&Mg\&Si  & 1.10 $\pm$ 0.02 &\cite{GREBENYUK20192546}\\
NUCLEON iron  & 1.17 $\pm$ 0.04 &\cite{GREBENYUK20192546}\\
CALET proton      & 1.01 $\pm$ 0.01 &\cite{CALET:2022vro}\\
CALET helium      & 1.01 $\pm$ 0.01 &\cite{CALET:2023nif}\\
CALET C\&O       & 1.09 $\pm$ 0.01 &\cite{Adriani:2020wyg}\\
CALET iron  & 1.07 $\pm$ 0.01 &\cite{CALET:2021fks}\\
\hline
\end{tabular}
\caption{Summary of energy calibration for space-born experiments.}
\label{tab:experiments}
\end{table}

\begin{table}[]
\centering
\begin{tabular}{ccc}
\hline
Experiment        & $\delta$ & Reference\\
\hline
HAWC            & 0.940 & \cite{HAWC:2022zma, HAWC:2021ubt}\\
LHAASO      & 0.955  &\cite{lhaaso2024allparticle}\\
TUNKA-133      & 0.955  &\cite{PROSIN201494, Prosin:2016rqu}\\
Auger        & 1.00  &\cite{PierreAuger:2021hun, PierreAuger:2013xim}\\
IceCube-IceTop      & 0.960  &\cite{IceCube:2019hmk}\\
IceTop   & 0.980  &\cite{IceCube:2020yct}\\
KASCADE  &  0.980 &\cite{Finger:2011bia}\\
KASCADE-Grande      & 1.00  &\cite{Apel:2013uni}\\
TA      & 0.945 &\cite{Ivanov:2020rqn, TelescopeArray:2018xyi}\\
TALE      & 0.99  &\cite{TelescopeArray:2018bya, TelescopeArray:2020bfv}\\
Tibet-III      & 1.00  &\cite{TIBETIII:2008qon}\\
\hline
\end{tabular}
\caption{Summary of energy calibration for ground-based experiments.}
\label{tab:ground}
\end{table}

\subsection{Spectrum Modeling\label{sec: spectrum}}
\subsubsection{Galactic components}

\textbf {A. contribution from a local source}\\

Precise measurements have revealed hardening in the energy spectra of protons and helium nuclei around hundreds of GV, with a subsequent softening near $\sim10$ TV~\cite{AMS:2021nhj, DAMPE:2019gys, Alemanno:2021gpb}, along with clear anisotropy patterns at these energies (see~\cite{Ahlers:2016rox} and references therein). Notably, the phase of dipole anisotropies experiences a reversal at $\sim100$ TeV, coinciding with the amplitude reaching its minimum. Below 100 TeV, the phase aligns roughly with the local interstellar magnetic field, as indicated by IBEX data~\cite{doi:10.1126/science.1245026}, while at higher energies, it shifts towards the Galactic center (GC) direction.

The effects of a local source could naturally account for both phenomena, as suggested in Refs~\cite{Liu:2018fjy, Fang:2020cru, Li:2021meq, Qiao:2022cge, Zhang:2021ipu, Zhang:2022pzt}.
The key point is that the phase reversal stems from the competition between GCR streamings: the background sources moving from the GC towards the anti-GC direction, dominating above 100 TeV, and the local source, from the source to its opposite, dominating below 100 TeV.

As previously outlined, the flux from the local source is phenomenologically described by a power-law spectrum with rigidity-dependent exponential cutoffs. Additionally, the propagation effect is taken into account that suppresses the local source's contribution at lower energies. Consequently, the contribution from the local source, denoted as $\Phi_{\mathrm{ls}}(E)$, is modeled as follows:
\begin{align}
\Phi_{\mathrm{ls}}(E)=\sum_{i} \phi_{\mathrm{ls}, i} \left(\frac{E}{\text{TeV}}\right)^{-\gamma_{\mathrm{ls}, i}} \exp \left(-\frac{E}{ZE_{c, \mathrm{ls}}}\right) \nonumber&\\
\times \exp \left(-\frac{L^2}{4D_{xx}\tau}\right)&\; ,
\end{align}
where $E$ represents the total energy, the sum extends over different primary particle types, $\phi_{\mathrm{ls}, i}$ denotes the normalization, $\gamma_{\mathrm{ls},i}$ the spectral index, $Z$ the atomic number, and $E_{c, \mathrm{ls}}$ the cutoff energy for proton. The local source's characteristics are defined by $L$ (distance to the source), $\tau$ (age of the source), and $D_{xx} = D_0(R/4\mathrm{GV})^\delta$ (the effective diffusion coefficient from the source to the Solar System).  Owing to the unknown nature and high degeneracy of these parameters, a set of representative values is selected and kept fixed throughout this study. These values are as follows: $L=500$ kpc, $\tau$=$2\times10^5$ year, $D_0$=$5\times10^{27}\;\mathrm{cm}^2\mathrm{s}^{-1}$, and $\delta$=0.33. 
\\

\textbf {B. soft and hard background components}\\

Different galactic CR components have already been considered in prior studies~\cite{Zatsepin:2006ci, Scrandis:2021hwu, Zhang:2022pzt,Gaisser:2013bla, Yue:2019sxt}. This necessity arises because the spectral index before the hardening at hundreds of GV is notably soft~\cite{AMS:2021nhj}. Assuming this to be the only background population, the all-particle spectrum would fall short of the measurements from ground-based experiments such as LHAASO~\cite{lhaaso2024allparticle} from hundreds of TeV to the knee. Therefore, a second background population with a harder spectrum is required.

The diversity of GCR sources may account for the multiple GCR components.
Possible GCR sources include different types of supernova remnants (SNRs) ~\cite{Hillas:2004nn, Hillas:2005cs}, stellar clusters~\cite{BINNS2008427, Aharonian:2018oau}, and Wolf-Rayet star explosions ~\cite{Thoudam:2016syr}. Additionally, CRs released at different SNR evolutionary phases could have different spectral indices and cutoff energies~\cite{Zhang:2017ksy, Zhang:2018fds}. Furthermore, CRs escaped into the Galactic halo could be re-accelerated by the Galactic wind termination shocks~\cite{1987ApJ/312/170J, 2006AdSpR/37/1923Z, Thoudam:2016syr}, potentially giving rise to a second population observed on Earth. It is noteworthy that propagation effects like spatial dependent deffusion~\cite{Tomassetti:2012ga, Guo:2015csa, Zhao:2021yzf} can also cause a hardening of the background spectrum, mimicking the effect of two background populations.

Given the above considerations, we introduce two background populations alongside the contributions from the local source. The flux from each population is represented by:
\begin{equation}
\Phi_{a}(E)=\sum_{i} \phi_{a, i} \left(\frac{E}{\text{TeV}}\right)^{-\gamma_{a, i}} \exp \left(-\frac{E}{ZE_{c, a}}\right)\; ,
\end{equation}
where $\Phi_{a}$ denotes the flux for population $a$ (Pop. I/Pop. II) at total energy $E$, the sum extends over different primary particle types within population $a$, $\phi_{a, j}$ represents the normalization, $\gamma_{a,i}$ the spectral index, $Z$ the atomic number, and $E_{c, a}$ the cutoff energy for proton. 

\vspace{2em}
In the model the predominant primaries include proton, helium, carbon, oxygen, nitrogen, magnesium, silicon, and iron. We also include a $Z=53$ group to Pop. II, potentially originating from r-process nucleosynthesis~\cite{Arnould:2007gh} as suggested in Ref.~\cite{Gaisser:2013bla}.
Other elements with lower abundances are neglected as they are not expected to have significant contributions to the total CR flux \cite{AMS:2021nhj,Engelmann:1990zz,Walsh:2020xon}. 

\subsubsection{Extragalactic components}
In the modeling of extragalactic contributions, in addition to the use of power-laws with rigidity-dependent exponential cutoffs, it is assumed that the extragalactic components are subject to exponential suppression at lower energies, consistent with the expectation of the `magnetic horizon effect’ \cite{Stanev:2000fb,Lemoine_2005, Berezinsky_2006, Mollerach_2013}. Following the approach in Ref.~\cite{Mollerach:2018lkt}, the extragalactic component is modeled as:
\begin{align}
\Phi_{\mathrm{xg}}(E)
=\sum_{i}\phi_{\mathrm{xg},i}\left(\frac{E}{100\mathrm{PeV}}\right)^{-\gamma_{\mathrm{xs}, i}} \exp \left(-\frac{E}{ZE_{c, \mathrm{xg}}}\right)&\\
\times\frac{1}{\cosh \left(ZE_l / E\right)}&\;,
\end{align}
where $E_l$ represents the characteristic energy scale below which the extragalactic contribution is suppressed. The $\cosh^{-1} \left(E_l / E\right)$ allows to implement a smooth suppression of the spectrum without affecting significantly the spectral shape above $2ZE_l$.

Identifying specific nuclei within the extragalactic components has always been challenging due to the limited statistics and complexities in mass composition reconstruction. In this study we build upon the latest findings from Pierre
Auger Collaboration~\cite{PierreAuger:2022atd,PierreAuger:2023htc,Salamida:2023qpx} (also reviewed in Refs.~\cite{Aloisio:2022xzy,Coleman:2022abf}), summarized as follows: above the ankle, the proportion of protons is remarkably low\footnote{The reason why proton cannot dominate above the knee can be understood through the following rationale. The mean logarithmic mass raise from $10^{18}$ eV to $10^{20}$ eV, eventually reaches $\langle\ln A\rangle\gtrsim2.0$.  If we hypothesize that protons constitute half of the spectrum, the other half would need to comprise heavier elements, like iron, to reproduce the $\langle\ln A\rangle$ measurements. However, such a composition would result in a significantly larger $\sigma(\ln A)$ which contradicts observations~\cite{PierreAuger:2013xim}}. Instead, the spectrum between the ankle and the cutoff is largely dominated by nuclei groups $2\leq A\leq 4$ and $5\leq A\leq 22$, as illustrated in figure 9 in Ref.~\cite{PierreAuger:2022atd} and figure 2 in Ref.~\cite{PierreAuger:2023htc}. Notably, the flux suppression observed at $E\sim4\times10^{19}$ eV is ascribed to rigidity cutoffs at the cosmic source, rather than being a result of propagation effects~\cite{Berezinsky:2007zza}.

Based on these findings, we assume the existence of two distinct extragalactic components. The HE component dominates the CR spectrum above the ankle and consists of helium and nitrogen, which correspond to the nuclear mass groups $2\leq A\leq 4$ and $5\leq A\leq 22$, respectively. Meanwhile, the LE component, contributing to the region between the knee and the ankle, consists of proton and helium, as suggested by Ref.~\cite{Aloisio:2013hya}.

\section{RESULTS \& DISCUSSION}\label{sec results}

\subsection{Spectra for individual species}

\begin{figure*}[htbp]
    \begin{center}
        \subfloat{\includegraphics[width=0.47\textwidth]{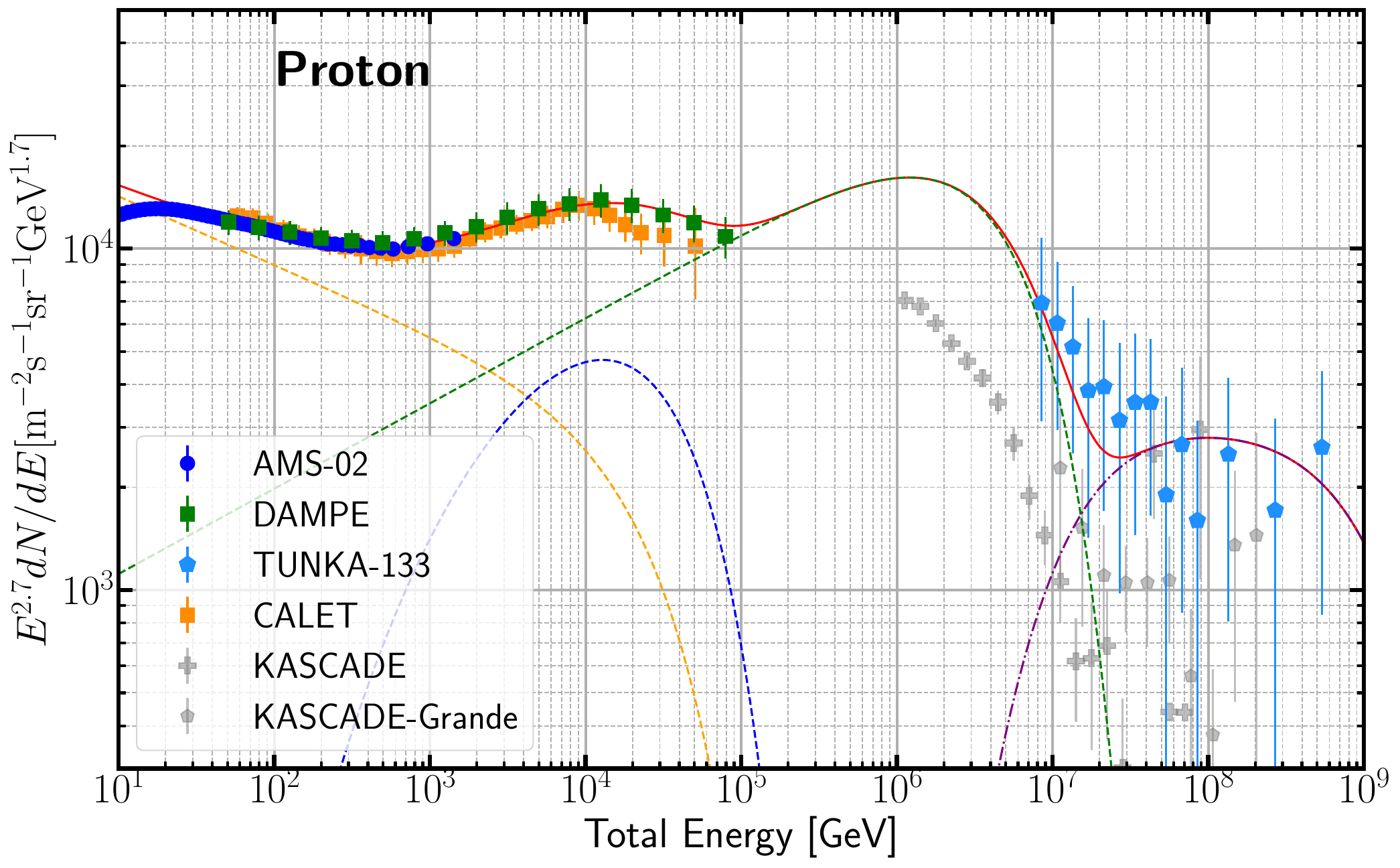}} \hskip 0.03\textwidth
        \subfloat{\includegraphics[width=0.47\textwidth]{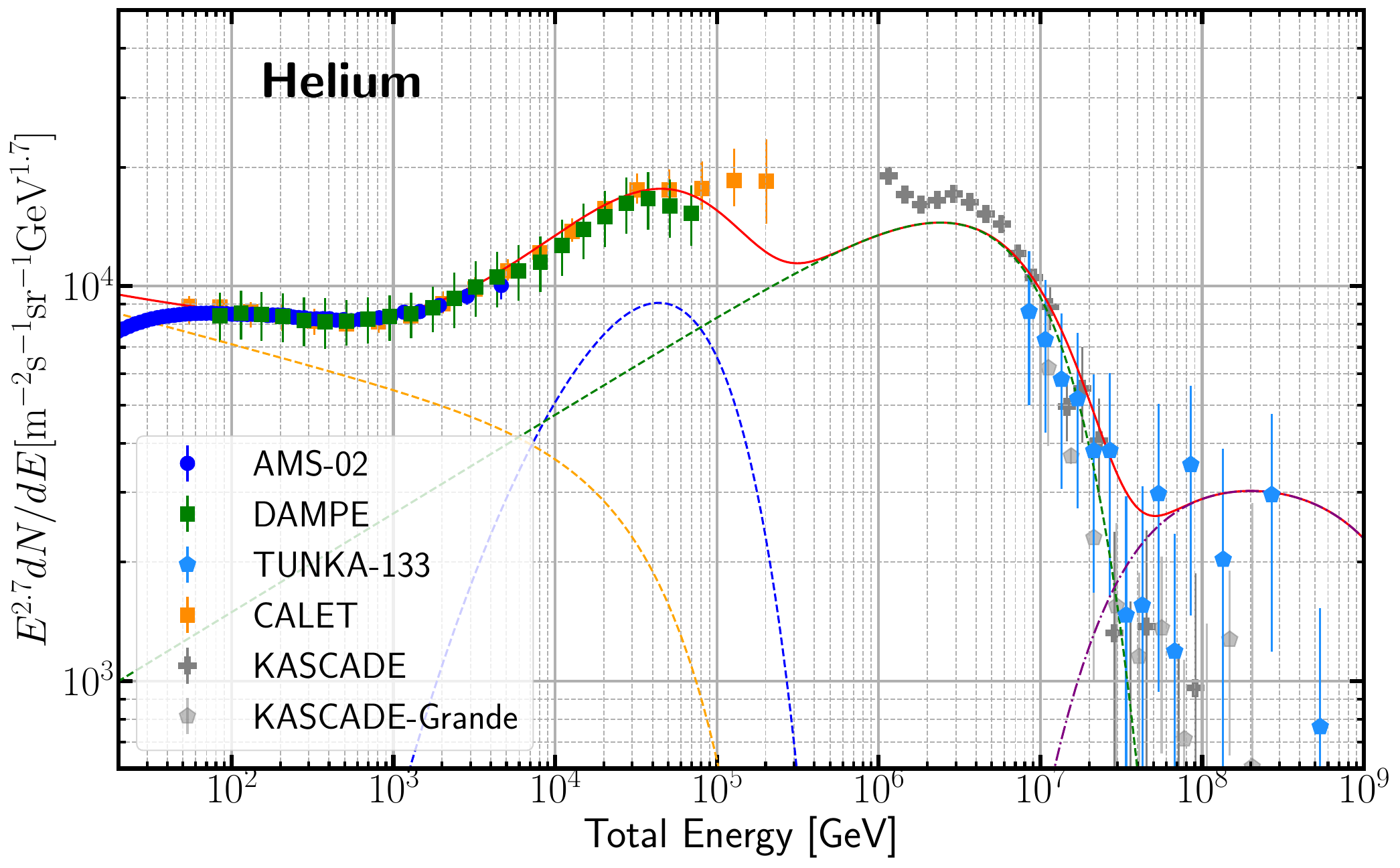}}
    \end{center}
    \begin{center}
        \subfloat{\includegraphics[width=0.47\textwidth]{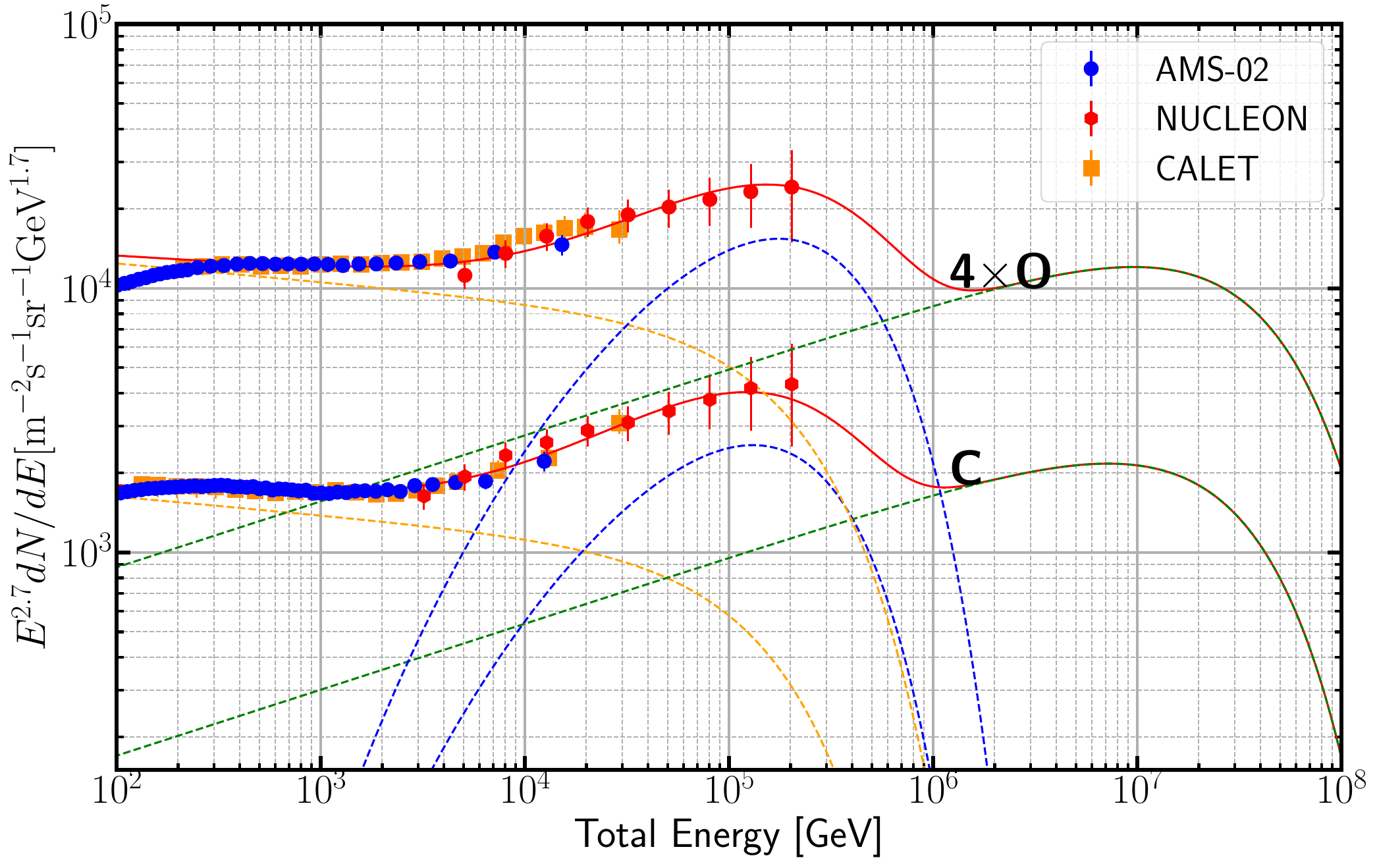}} \hskip 0.03\textwidth
        \subfloat{\includegraphics[width=0.47\textwidth]{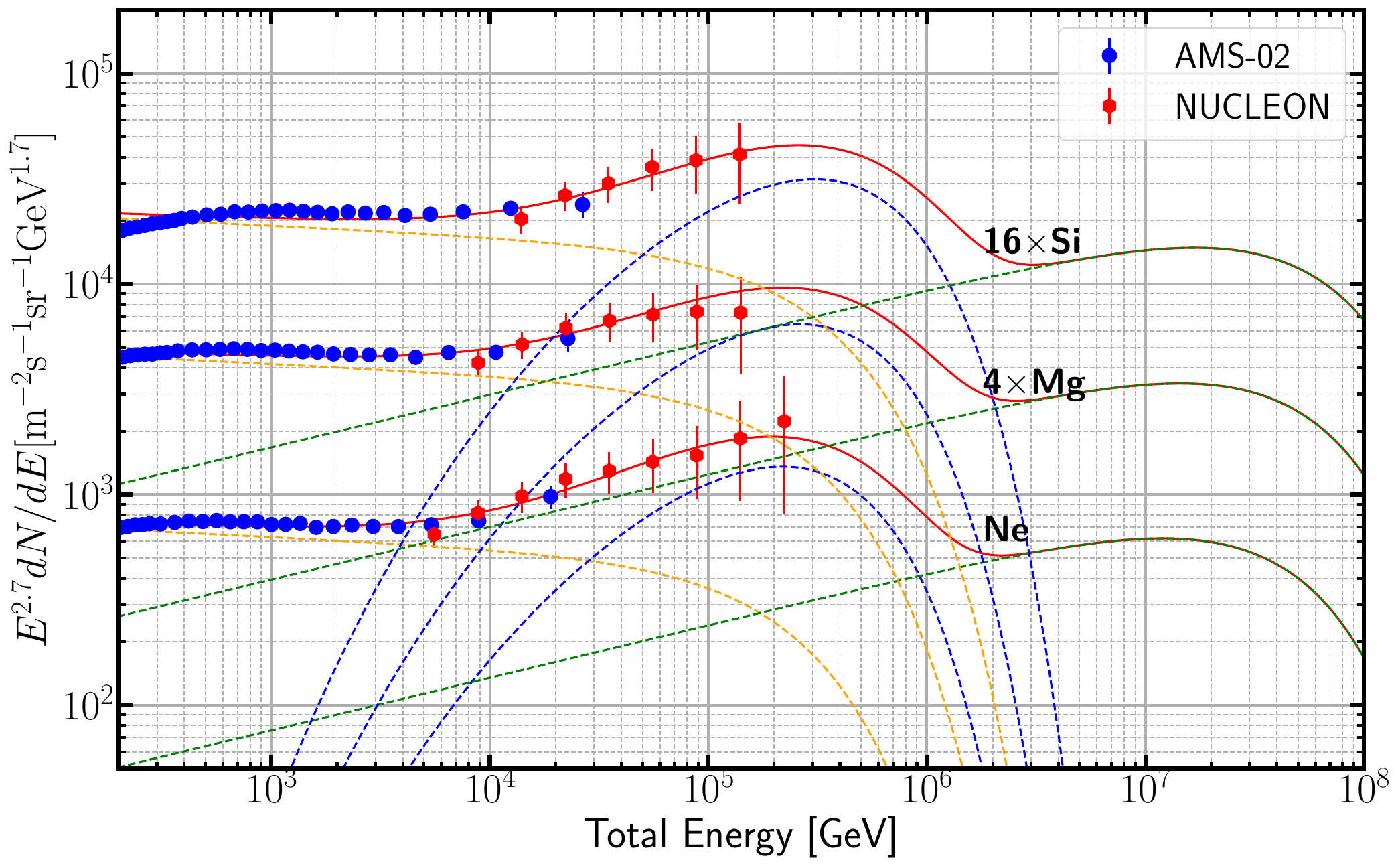}}
    \end{center}
    \begin{center}
        \subfloat{\includegraphics[width=0.47\textwidth]{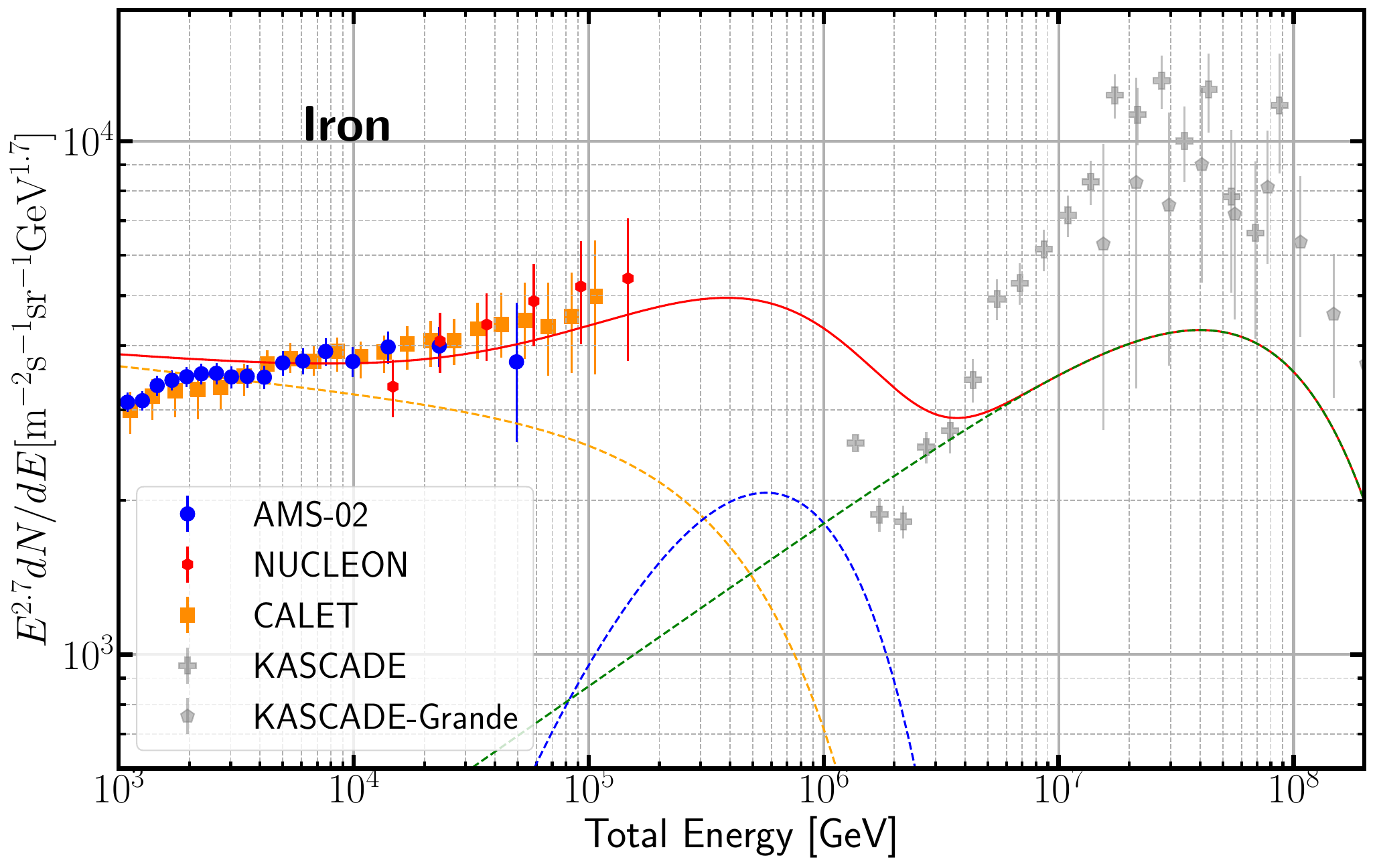}} \hskip 0.03\textwidth
        \subfloat{\includegraphics[width=0.47\textwidth]{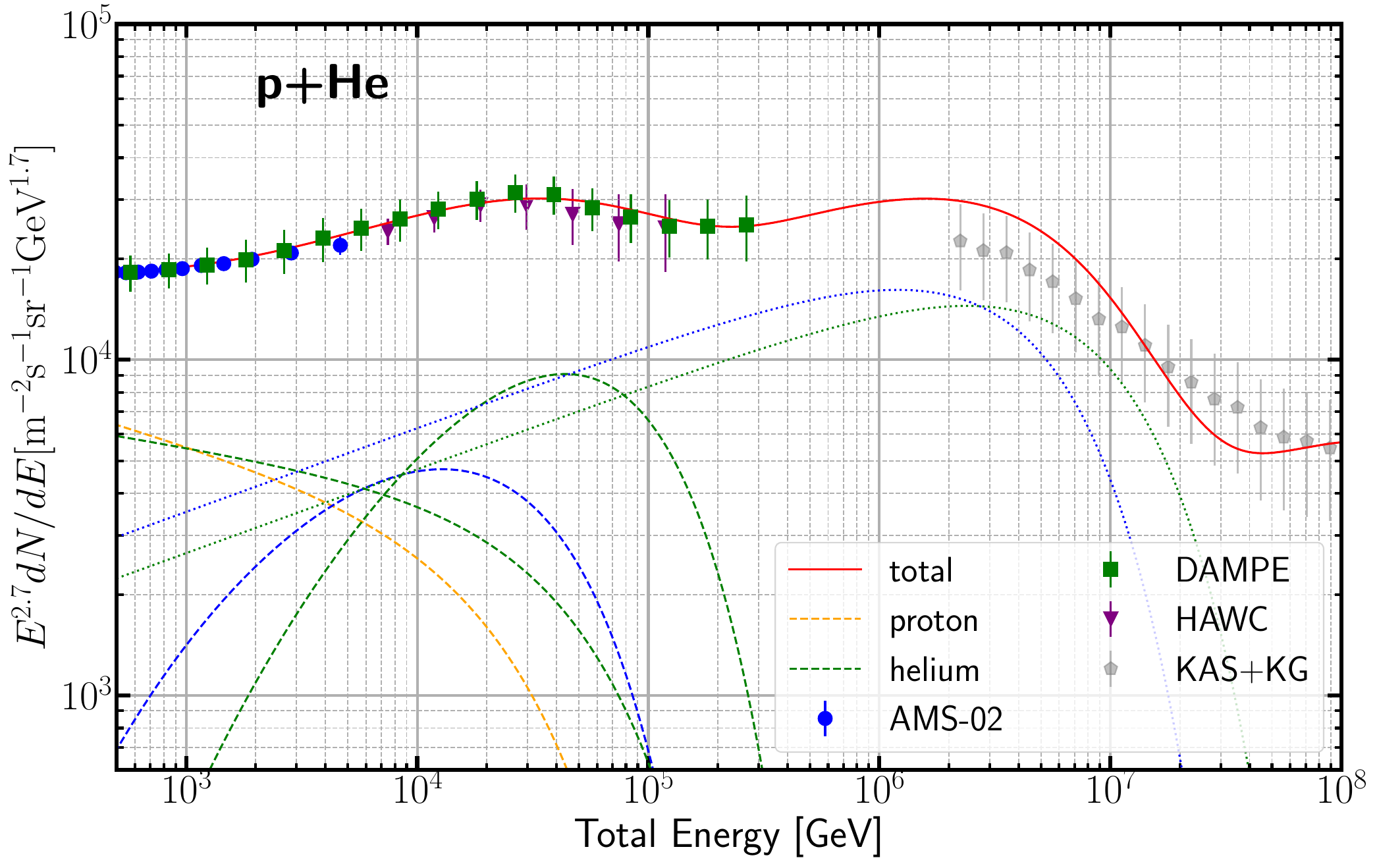}}
    \end{center}
    \begin{center}
        \subfloat{\includegraphics[width=0.47\textwidth]{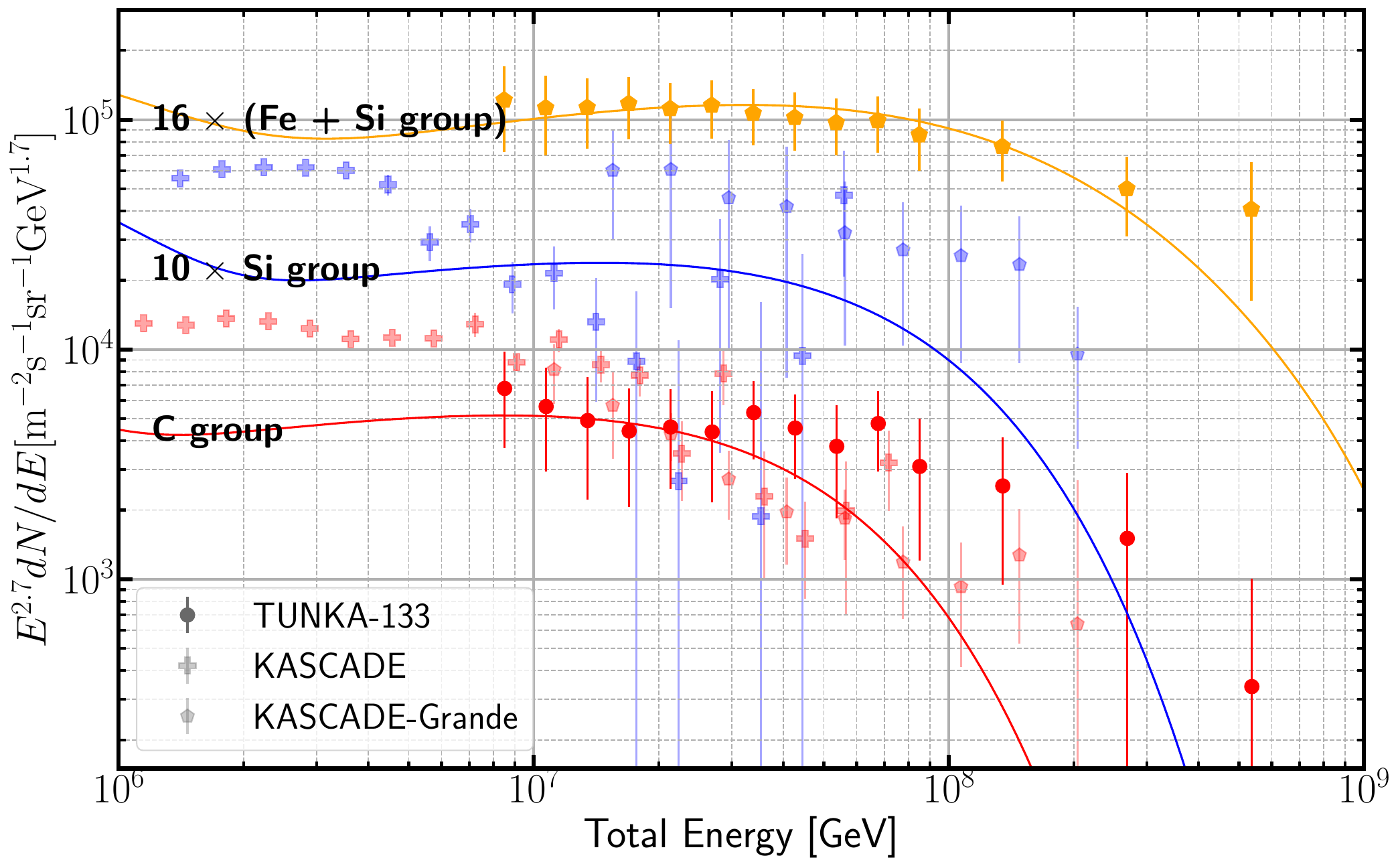}} \hskip 0.03\textwidth
        \subfloat{\includegraphics[width=0.47\textwidth]{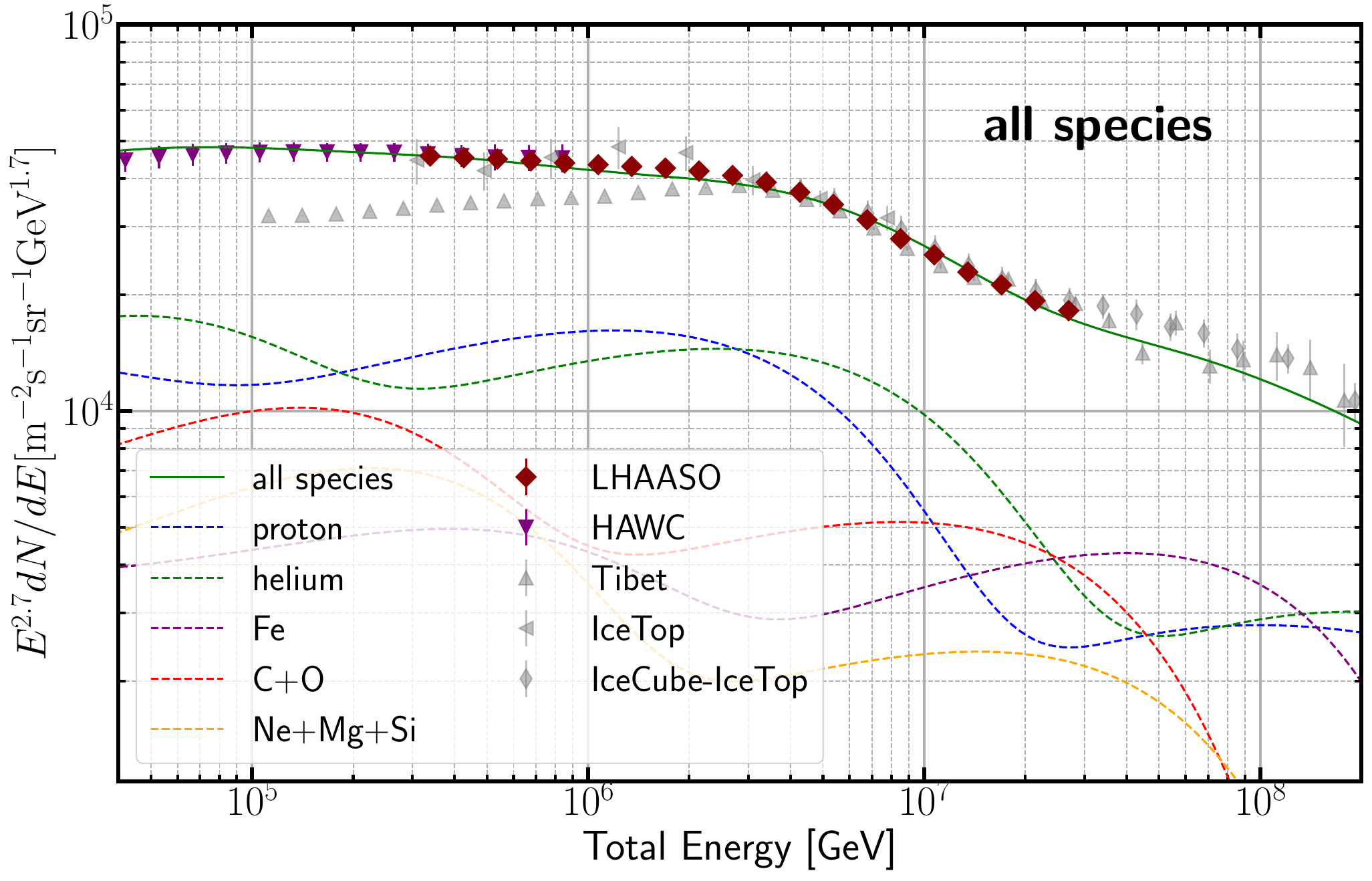}}
    \end{center}
    \captionsetup{justification=raggedright}
\end{figure*}
\begin{figure*}[t!]
    \caption{Fitted energy spectra of our model, compared with the data. In each panel, the orange, blue, and green dashed lines show the contributions of Pop. I, local source and Pop. II, respectively, while the solid lines are the total contribution. The grey data points are not used in the model building. References of the data: protons, AMS-02~\cite{AMS:2021nhj}, DAMPE~\cite{DAMPE:2019gys}, TUNKA-133~\cite{PROSIN201494}, CALET~\cite{CALET:2022vro} KASCADE~\cite{Finger:2011bia}, KASCADE-Grande~\cite{Apel:2013uni}; Helium, AMS-02~\cite{AMS:2021nhj}, DAMPE~\cite{Alemanno:2021gpb}, TUNKA-133~\cite{PROSIN201494}, CALET~\cite{CALET:2023nif} KASCADE~\cite{Finger:2011bia}, KASCADE-Grande~\cite{Apel:2013uni}; Carbon and Oxygen, AMS-02~\cite{AMS:2021nhj}, NUCLEON~\cite{GREBENYUK20192546}, CALET~\cite{Adriani:2020wyg}; Neon, Magnesium, and Silicon, AMS-02~\cite{AMS:2021nhj}, NUCLEON~\cite{GREBENYUK20192546}; Iron, AMS-02~\cite{AMS:2021nhj}, CALET~\cite{CALET:2021fks} KASCADE~\cite{Finger:2011bia}, KASCADE-Grande~\cite{Apel:2013uni}; p+He, AMS-02~\cite{AMS:2021nhj}, DAMPE~\cite{dampecollaboration2023measurement}, 
    HAWC~\cite{HAWC:2022zma}, KAS+KG~\cite{KASCADE-Grande:2015svq}; groups, TUNKA-133~\cite{PROSIN201494}, KASCADE~\cite{Finger:2011bia}, KASCADE-Grande~\cite{Apel:2013uni}; all-particle, LHAASO~\cite{lhaaso2024allparticle}, HAWC~\cite{HAWC:2021ubt}, Tibet~\cite{TIBETIII:2008qon}, IceTop~\cite{IceCube:2020yct}.}\label{fig:results}
\end{figure*}
\begin{table*}[]
\captionsetup{justification=raggedright}
\caption{Spectral parameters of the galactic components.}
\begin{tabular}{cccccccccc}
\hline
Species & \multicolumn{2}{c}{Pop. I: $E_c = 30$ TV} & \multicolumn{2}{c}{Loc. Source: $E_c = 36$ TV} & \multicolumn{2}{c}{Pop. II: $E_c = 4.8$ PV}\\
 & $\Phi_{0,i}$ & $\gamma_i$ &  $\Phi_{0,i}$ & $\gamma_i$ & $\Phi_{0,i}$ & $\gamma_i$ \\
 & (m$^{-2}$s$^{-1}$sr$^{-1}$GeV$^{-1}$) & &  (m$^{-2}$s$^{-1}$sr$^{-1}$GeV$^{-1}$) & &  (m$^{-2}$s$^{-1}$sr$^{-1}$GeV$^{-1}$) & &\\
\hline
p  & $4.5 \times 10^{-5}$ & 2.90 & $2.3 \times 10^{-4}$ & 2.77 & $2.8 \times 10^{-5}$ & 2.45\\
He & $4.4 \times 10^{-5}$ & 2.81 & $1.5 \times 10^{-4}$ & 2.45 & $2.1 \times 10^{-5}$ & 2.45 \\
C  & $1.1 \times 10^{-5}$ & 2.77 & $3.2 \times 10^{-5}$ & 2.45 & $2.4 \times 10^{-6}$ & 2.45 \\
O  & $2.1 \times 10^{-5}$ & 2.77 & $4.5 \times 10^{-5}$ & 2.45 & $3.1 \times 10^{-6}$ & 2.45 \\
Ne & $5.0 \times 10^{-6}$ & 2.75 & $1.5 \times 10^{-5}$ & 2.45 & $6.0 \times 10^{-7}$ & 2.45 \\
Mg & $8.3 \times 10^{-6}$ & 2.75 & $1.7 \times 10^{-5}$ & 2.45 & $7.8 \times 10^{-7}$ & 2.45 \\
Si & $9.4 \times 10^{-6}$ & 2.75 & $2.0 \times 10^{-5}$ & 2.45 & $8.3 \times 10^{-7}$ & 2.45 \\
Fe & $2.9 \times 10^{-5}$ & 2.75 & $1.8 \times 10^{-5}$ & 2.45 & $1.6 \times 10^{-6}$ & 2.38 \\
Z=53 & & & & &$1.5 \times 10^{-8}$ & 2.10 \\
\hline
\end{tabular}
\label{table:galactic}
\end{table*}

\begin{table*}[]
\captionsetup{justification=raggedright}
\caption{Spectral parameters of the galactic components.}
\begin{tabular}{ccccc}
\hline
Species & \multicolumn{2}{c}{LE: $E_c = 1$ EV, $E_l = 19$ PV}\; & \multicolumn{2}{c}{HE: $E_c = 4$ EV, $E_l = 1.5$ EV} \\
& $\Phi_{0,i}$ & $\gamma_i$ &  $\Phi_{0,i}$ & $\gamma_i$\\
 & (m$^{-2}$s$^{-1}$sr$^{-1}$GeV$^{-1}$) & &  (m$^{-2}$s$^{-1}$sr$^{-1}$GeV$^{-1}$)\\
\hline
p  & $8.0 \times 10^{-19}$ & 2.67 &  \\
He & $8.5 \times 10^{-19}$ & 2.67 & $3.1 \times 10^{-20}$ & 2.0\\
N  & & & $1.7 \times 10^{-20}$ & 2.0\\
\hline
\end{tabular}
\label{table:extragalactic}
\end{table*}

The individual spectrum of the major species, alongside the all-particle spectrum below the second-knee, are shown in Figure~\ref{fig:results}. Data points used in the model building, represented as colored solid points, and additional datasets not used, illustrated as grey transparent points, are both presented. The spectral parameters defining both the Galactic and extragalactic components are summarized in Table~\ref{table:galactic} and Table~\ref{table:extragalactic}, respectively. Our model successfully reproduces every individual spectrum used in the model-building process. 

The three Galactic components in our model play distinct roles in shaping the primary CR spectrum. Pop. I, characterized by a soft spectrum with a cutoff at 30 TV, predominantly influences the spectrum up to the hardening observed at approximately 200 GV. The local source, with a cutoff at 36 TV, is primarily responsible for the bump observed in the proton and helium spectra around 10 TV. 
Pop. II, marked by a cutoff at 4.8 PV and a spectrum harder than that of Pop. I  contributes to the formation of the knee. Regarding the extragalactic components, the LE component, with a cutoff at 1 EV and a suppression below 19 PV, influences the spectrum from the low-energy ankle to the ankle. Conversely, the HE component, which cuts at 4 EV and attenuates below 1.5 EV, dominates the spectrum above the ankle.

The datasets not used in our model building, for the most part, are compatible with our model when considering systematic uncertainties. Notably, the KASCADE experiment and its extension, KASCADE-Grande, provide comprehensive results for different CR element groups. Our model reproduces most of their findings, given the moderate differences between these two experiments. However, the iron spectrum might be an exception, where our model's predictions systematically fall below both sets of measurements above $\sim10^7$ GeV. This deviation is further illustrated in the lower panel of Figure~\ref{fig:results_high}, where the predicted $\langle\ln A\rangle$ is lower than that measured by KASCADE and KASCADE-Grande. To reconcile our model with the KASCADE and KASCADE-Grande measurements, a harder Pop. II iron spectrum would be necessary, adjusting from the current one ($\gamma=2.38$) to $\sim2.3$. As for the other experiments, measurements from IceCube/IceTop agree well with our model prediction, whereas data from Tibet-III suggest the need for a generally harder Pop. II spectral index.

\subsection{All-particle spectrum}

\begin{figure}[htbp]
\includegraphics[width=0.48\textwidth]{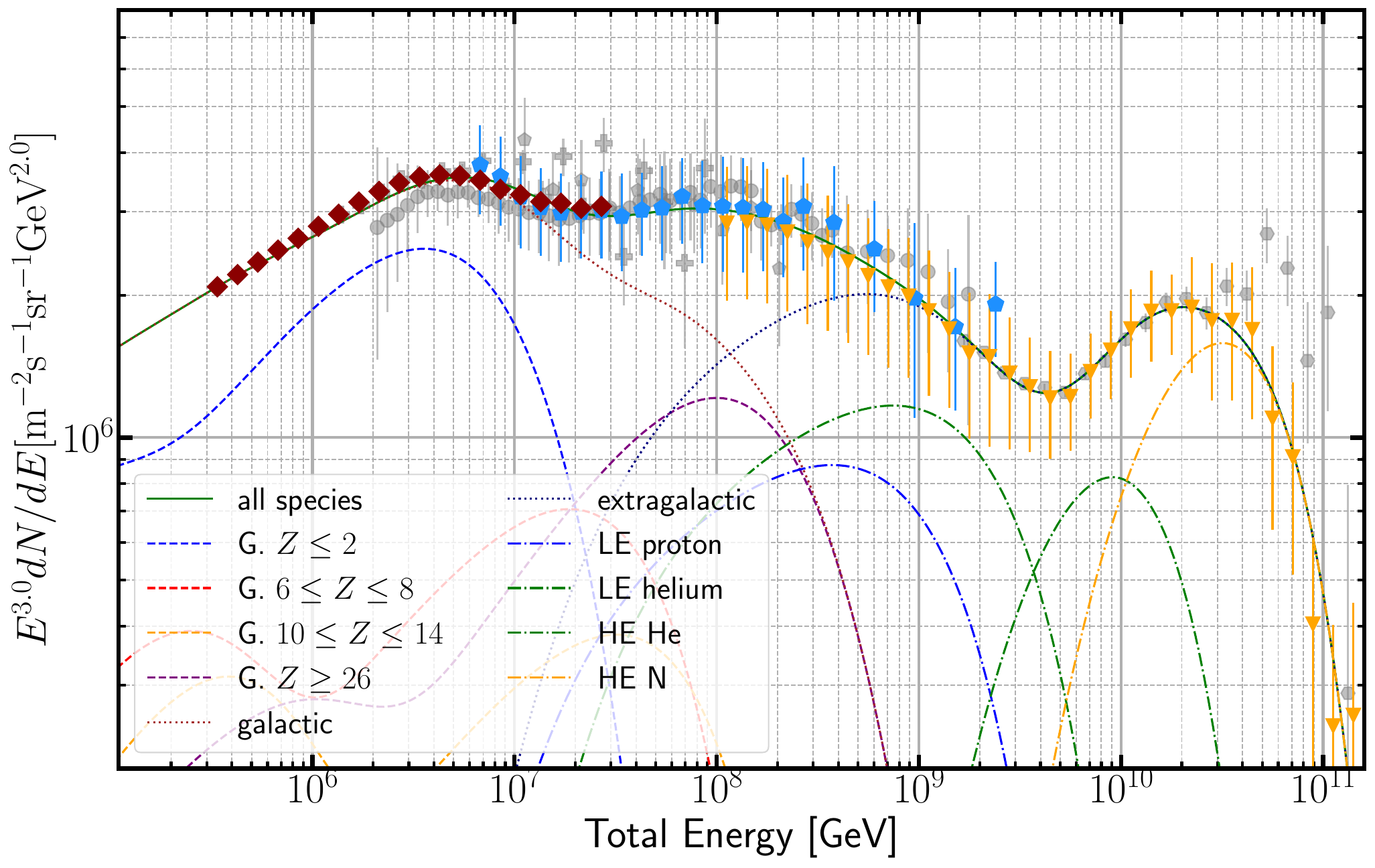}\\
\includegraphics[width=0.48\textwidth]{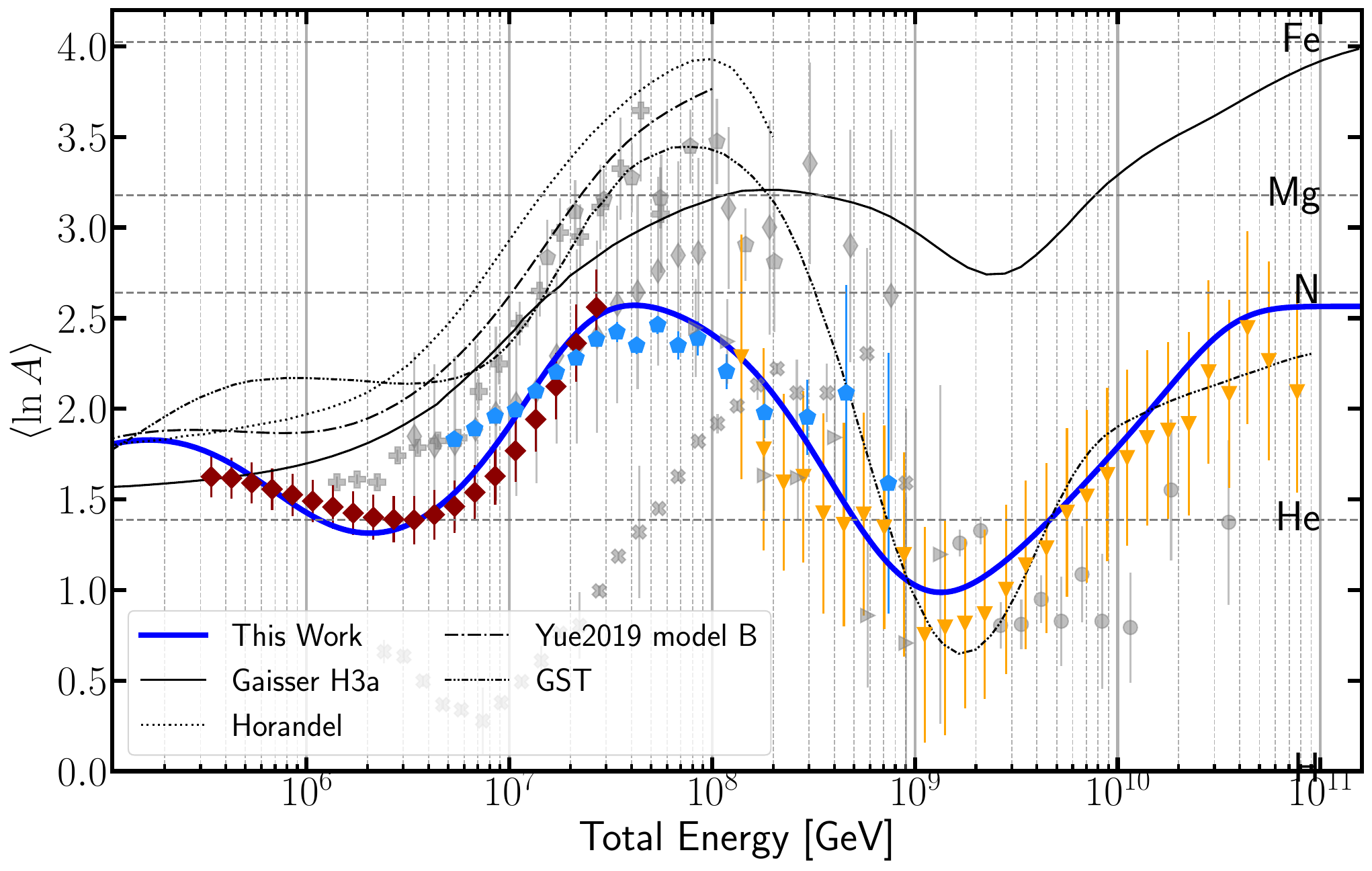}\\
\includegraphics[width=0.48\textwidth]{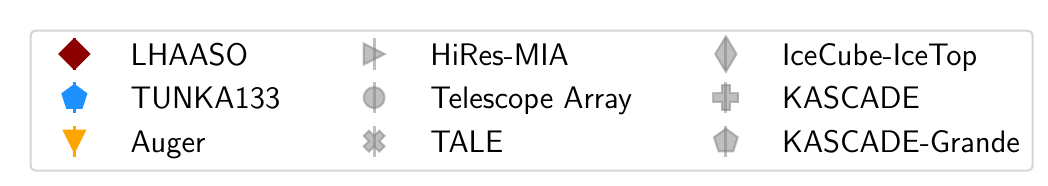}\\
\captionsetup{justification=raggedright}
    \caption{All-particle spectrum (upper panel) and the mean logarithmic mass (lower panel) from $2\times 10^{14}$ eV to $2\times10^{11}$ eV. `G' stands for `galactic' in the upper panel. References: all-particle, LHAASO~\cite{lhaaso2024allparticle}, TUNKA-133~\cite{PROSIN201494}, Auger~\cite{PierreAuger:2021hun}, TALE~\cite{TelescopeArray:2018bya}, TA~\cite{Ivanov:2020rqn},
    KASCADE~\cite{Finger:2011bia}, KASCADE-Grande~\cite{Apel:2013uni}; $\langle \ln A\rangle$, LHAASO~\cite{lhaaso2024allparticle}, TUNKA-133~\cite{Prosin:2016rqu}, Auger~\cite{PierreAuger:2013xim}, HiRes-MIA~\cite{Kampert:2012mx}, TALE~\cite{TelescopeArray:2020bfv}, Telescope Array~\cite{TelescopeArray:2018xyi}, 
    IceCube-IceTop~\cite{IceCube:2019hmk}, KASCADE~\cite{Finger:2011bia}, KASCADE-Grande~\cite{Apel:2013uni}, Gaisser H3a~\cite{Gaisser:2013bla}, Horandel~\cite{Hoerandel:2002yg}, Yue2019 model B~\cite{Yue:2019sxt}, GST~\cite{Stanev:2014mla}. The other references are the same as in Figure~\ref{fig:results}.}\label{fig:results_high}
\end{figure}

The all-particle spectrum predicted by our model is illustrated in the upper panel of Figure~\ref{fig:results_high}. Our model accurately captures the major spectral features in the all-particle spectrum and the evolution of mass composition with energy. Data not included in the model-building process are represented as transparent grey points, most of which  are successfully reproduced by our model, although there are a few deviations that will be addressed subsequently.

The transition to extragalactic CRs in our model happens just above $10^{17}$ eV, in agreement with the findings in Refs.~\cite{Berezhko:2007gh, Aloisio:2013hya, Sveshnikova:2014yes, Giacinti:2015hva, Mollerach:2018lkt}. However, there are suggestions that the extragalactic LE component, consisting of light elements, may have a Galactic origin, such as Wolf-Rayet stars as proposed in Ref.~\cite{Thoudam:2016syr}. In this scenario the transition is close to the ankle, consistent with earlier beliefs~\cite{Rachen:1993gf, Wibig:2004ye, Berezinsky:2007zza}. Nonetheless, the low level of observed anisotropy in the energy ranges $10^{17}$ -- $10^{18}$ eV~\cite{PierreAuger:2020fbi} disfavors such a Galactic origin scenario ~\cite{Giacinti:2011ww,PierreAuger:2012gro}\footnote{The `population B' proposed by Hillas~\cite{Hillas:2004nn,Hillas:2005cs}, characterized by a large iron fraction, could be compatible with the anisotropy constraints. Yet, this scenario is strongly disfavored by composition measurements from the Auger Collaboration ~\cite{PierreAuger:2014sui, PierreAuger:2014gko}, as noted in Ref.~\cite{Giacinti:2015hva}.}.

Next, we turn to the formation of various spectral features in our model. The spectral knee is attributed to the cutoff in the proton spectrum of Pop. II at 4.8 PeV. This feature is further broadened due to the cutoff in the helium spectrum, occurring at 9.6 PeV. Our findings align with those presented in Refs.~\cite{Guo:2017tle, Dembinski:2017zsh, Mollerach:2018lkt}, offering a contrast to earlier studies which suggested that the knee is dominated by nuclei heavier than helium~\cite{Shibata:2010zza,Zhao:2015rja,Thoudam:2016syr}. The reason for a light knee originates from its mass composition, as measured by LHAASO~\cite{lhaaso2024allparticle}, where the mean logarithmic mass is heavier than He and lighter than CNO, suggesting that the first cut-off of the all-particle energy spectrum is attributed to lighter elements, rather than to medium-heavy elements. 

The spectral hardening around 20 PeV, known as the low-energy ankle, is attributed to the increasing contribution of the extragalactic proton and helium, as supported by recent studies~\cite{Sveshnikova:2014yes, Mollerach:2018lkt}. Specifically, the total Galactic contribution exhibits no significant hardening, as is apparent from the upper panel of Figure~\ref{fig:results_high}. This interpretation contrasts with earlier theories, such as those proposed in Ref.~\cite{Sveshnikova:2013qxa}, in which the hardening was due to the rise of heavy galactic components contributing to the knee. The rationale for favoring an extragalactic origin is twofold. Firstly, achieving such hardening would necessitate a substantially harder spectral index for medium-heavy elements compared to protons and helium, which is physically implausible at such high energies where the rest mass is negligible. Secondly, even if the spectral feature is reproduced, the resulting mean logarithmic mass around 20 PeV would be too high compared with LHAASO and TUNKA-133 measurements.

The second-knee occurs at around 200 PeV in our model and is attributed to the steepening of the Galactic iron and elements in the $Z=53$ group. 
This slightly contrasts with the findings from Refs.\cite{Mollerach:2018lkt, Abu-Zayyad:2018btv}, which attribute the second-knee exclusively to the steepening of Galactic iron. The discrepancy between these interpretations and ours stems from the choice of datasets: Refs.\cite{Mollerach:2018lkt, Abu-Zayyad:2018btv} rely on TALE data that locates the second-knee at 100 PeV, whereas our analysis is based on Auger and TUNKA-133 data, which place the second-knee at 200 PeV. Consequently, the ratio $E_{\mathrm{2^{\mathrm{nd}} knee}}/E_{\mathrm{knee}}$ changes from $\sim26$ (the atomic number of iron) in the former scenario to $\sim53$ in the latter. Given these systematic uncertainties, reaching a definitive conclusion remains challenging.

Regarding the spectral region at and beyond the ankle, our model aligns with the findings of the Auger Collaboration~\cite{PierreAuger:2022atd,PierreAuger:2023htc,Salamida:2023qpx}. Specifically, the helium and CNO nuclei, characterized by a notably hard spectrum\footnote{In our model, the spectral index for the HE component is set at -2.0, in contrast to the Auger Collaboration's findings, which suggest even harder indices, exceeding 1.0~\cite{PierreAuger:2022atd}. This divergence arises because our model includes a low energy suppression for the HE component at 4 EV, resulting in a notably steep overall spectrum.} and a rigidity-dependent cutoff, are responsible for the formation of the ankle at $\sim5$ EeV and the subsequent cutoff at $\sim40$ EeV.  
The all-particle spectrum measured by TA is higher than that observed by Auger at the highest energies, a discrepancy confirmed by the joint Auger-TA working group~\cite{Deligny:2020gzq, PierreAuger:2023wti}. However, this discrepancy arises at such high energies that it does not impact the main conclusions of this study. We refer the reader to Ref.~\cite{Plotko:2022urd} and references therein for discussion on possible reasons behind this mismatch.

{\subsection{The mean logarithmic mass}}

As Hillas has demonstrated, ``The smoothness of the total spectrum can hide large bumps in individual components"~\cite{Hillas:2006ms}. Consequently, the crucial insights provided by the mean logarithmic mass are essential for the construction of our model. 
Our model's prediction of the mean logarithmic mass is shown in the lower panel of Figure~\ref{fig:results_high}, along with the prediction of previous models~\cite{Hoerandel:2002yg, Gaisser:2013bla, Stanev:2014mla, Dembinski:2017zsh, Yue:2019sxt}. 

It can be seen in Fig. \ref{fig:results_high} the decreases in the mean logarithmic mass -- the first decrease spanning 0.3 to 3 PeV, followed by a second decrease ranging from $\sim40$ PeV to $\sim1$ EeV -- stem from similar origins. The cutoff in the local iron source naturally accounts for the first decrease in the mean logarithmic mass. This trend is  evident in the lower left panel of Figure~\ref{fig:results}. The previous models~\cite{Hoerandel:2002yg, Gaisser:2013bla, Stanev:2014mla, Yue:2019sxt} fail to depict this decrease, which demonstrates the importance of the galactic local contribution included in our model. The second decrease in the mean logarithmic mass is associated with the cutoff of iron and $Z=53$ group elements in the Pop. II component. In summary, both decreases in the average logarithmic mass result from the cutoff of the heavy elements, the former from the local source and the latter from the Pop. II component.

Likewise, the causes for the two observed rises in the mean logarithmic mass, with the first spanning from 0.3 PeV to 40 PeV and the second starting at 1 EeV, are consistent. Both rises are primarily attributed to the cutoff of the proton contributions, the former from the Galactic component and the latter from the LE extragalactic component.

We notice the correlation between features in the energy spectrum and the mean logarithmic mass. The local source, which accounts for the 10 TV bump in the proton and helium spectrum, naturally leads to the observed decrease from 0.3 to 3 PeV, as measured by LHAASO. Similarly, the cutoff in the galactic components, resulting in the knee, naturally accounts for the decrease observed by Auger. 
Therefore the features in the spectrum and in the mean logarithmic mass are natural consequences of its multi-component nature. 

It should be noted that there are large discrepancies among different experiments' measurements of $\langle\ln A\rangle$.
At the higher end of the energy spectrum, the TA results are consistent with those from Auger within statistical and systematic errors, as shown by the joint Auger-TA working group~\cite{PierreAuger:2023yym}. A fit to the TA data indicates a predominance of protons at the highest energies~\cite{Bergman:2021djm}. Consequently, this suggests that our model's HE nitrogen component should be replaced by a proton component. 
The results from KASCADE and IceCube-IceTop may require a harder Pop. II iron component.
\\

\section{SUMMARY\label{sec:conclusion}}

The LHAASO Collaboration's measurements on the all-particle spectrum and the mean logarithmic mass with unprecedented precision are extremely important for understanding the origins of both Galactic and extragalactic cosmic rays. 
%
In this study, we develop a phenomenological model that accurately reproduces all individual spectral measurements obtained from space-borne experiments, alongside the all-particle spectrum and the mean logarithmic mass reported by ground-based experiments, covering the entire energy range between tens of $\mathrm{GeV}/Z$ to $10^{11}$ GeV. The different spectral features and various changes in  $\langle\ln A\rangle$ are naturally explained by the alternation contributions from different components. The main results obtained through this study indicate:
\begin{itemize}
  \item To explain the CR spectrum and composition, a minimum of three Galactic components are required: two background populations, one with a soft and the other with a harder spectral index, and a local source. Additionally, two extragalactic components are required.
  \item The spectral bump observed in the proton and helium spectrum around 10 TV originates from a local source's influence.
  \item The presence of the knee at approximately 4 PeV is attributed to the cutoff of galactic proton and helium contributions, indicating a light knee.
  \item The emergence of the low-energy ankle near 20 PeV stems from the increasing contributions from the extragalactic proton and helium.
  \item The formation of the second-knee around 200 PeV is linked to the cutoff of the galactic iron and potentially the elements within the $Z=53$ group. The transition to extragalactic cosmic rays happens just before the second-knee.
  \item The ankle hardening at $\sim5$ EeV is a purely extragalactic phenomenon, specifically, the contribution of an extremely hard HE extragalactic component composed of helium and CNO. The rigidity-dependent cutoff of this component results in the cutoff
  at $\sim40$ EeV.
  \item Variations in the mean logarithmic mass share common causes: the decreases result from the cutoff of the heavy elements, initially by the local source followed by the galactic component, whereas the rises are related to the cutoff of the proton contributions, initially from the galactic and then from the LE extragalactic component.
\end{itemize}

Upcoming measurements by LHHAASO focusing on various CR groups (p, He, C, Si, Fe)~\cite{Zhang:2023gbv}, with larger statistics and reduced systematic errors, will serve as a critical evaluation of our model. Our model distinctly predicts an iron knee near 500 TeV and reduced iron flux relative to that observed by KASCADE/KASCADE-Grande~\cite{Finger:2011bia, Apel:2013uni} beyond roughly 10 PeV.

Future work will build on these findings to refine our understanding of cosmic ray origins and propagation. Ongoing developments in CR observational technology, such as High Energy cosmic-Radiation Detector (HERD)~\cite{Cattaneo:2019uui}, Probe of Extreme Multi-Messenger Astrophysics (POEMMA)~\cite{POEMMA:2020ykm},  NeUtrino and Seismic Electromagnetic Signals space mission (NUSES)~\cite{DiSanto:2023lcx}, IceCube-Gen2~\cite{IceCube-Gen2:2020qha}, Fluorescence detector Array of Single-pixel Telescopes (FAST)~\cite{Fujii:2015dra}, Giant Radio Array for Neutrino Detection (GRAND)~\cite{deMelloNeto:2023zvk}, Global Cosmic-ray Observatory (GCOS)~\cite{AlvesBatista:2023lqg}, and others, alongside advances in theoretical modeling, are set to deepen our understanding of cosmic rays.

\acknowledgments
This work is supported by 
the National Natural Science Foundation of China under
the Grants No. 12175248 and No. 12105292. 

\bibliography{apssamp}

\begin{thebibliography}{119}%
\makeatletter
\providecommand \@ifxundefined [1]{%
 \@ifx{#1\undefined}
}%
\providecommand \@ifnum [1]{%
 \ifnum #1\expandafter \@firstoftwo
 \else \expandafter \@secondoftwo
 \fi
}%
\providecommand \@ifx [1]{%
 \ifx #1\expandafter \@firstoftwo
 \else \expandafter \@secondoftwo
 \fi
}%
\providecommand \natexlab [1]{#1}%
\providecommand \enquote  [1]{``#1''}%
\providecommand \bibnamefont  [1]{#1}%
\providecommand \bibfnamefont [1]{#1}%
\providecommand \citenamefont [1]{#1}%
\providecommand \href@noop [0]{\@secondoftwo}%
\providecommand \href [0]{\begingroup \@sanitize@url \@href}%
\providecommand \@href[1]{\@@startlink{#1}\@@href}%
\providecommand \@@href[1]{\endgroup#1\@@endlink}%
\providecommand \@sanitize@url [0]{\catcode `\\12\catcode `\$12\catcode
  `\&12\catcode `\#12\catcode `\^12\catcode `\_12\catcode `\%12\relax}%
\providecommand \@@startlink[1]{}%
\providecommand \@@endlink[0]{}%
\providecommand \url  [0]{\begingroup\@sanitize@url \@url }%
\providecommand \@url [1]{\endgroup\@href {#1}{\urlprefix }}%
\providecommand \urlprefix  [0]{URL }%
\providecommand \Eprint [0]{\href }%
\providecommand \doibase [0]{https://doi.org/}%
\providecommand \selectlanguage [0]{\@gobble}%
\providecommand \bibinfo  [0]{\@secondoftwo}%
\providecommand \bibfield  [0]{\@secondoftwo}%
\providecommand \translation [1]{[#1]}%
\providecommand \BibitemOpen [0]{}%
\providecommand \bibitemStop [0]{}%
\providecommand \bibitemNoStop [0]{.\EOS\space}%
\providecommand \EOS [0]{\spacefactor3000\relax}%
\providecommand \BibitemShut  [1]{\csname bibitem#1\endcsname}%
\let\auto@bib@innerbib\@empty
\bibitem [{\citenamefont {Hess}(1912)}]{Hess:1912srp}%
  \BibitemOpen
  \bibfield  {author} {\bibinfo {author} {\bibfnamefont {V.~F.}\ \bibnamefont
  {Hess}},\ }\href@noop {} {\bibfield  {journal} {\bibinfo  {journal} {Phys.
  Z.}\ }\textbf {\bibinfo {volume} {13}},\ \bibinfo {pages} {1084} (\bibinfo
  {year} {1912})}\BibitemShut {NoStop}%
\bibitem [{\citenamefont {Becker~Tjus}\ and\ \citenamefont
  {Merten}(2020)}]{BeckerTjus:2020xzg}%
  \BibitemOpen
  \bibfield  {author} {\bibinfo {author} {\bibfnamefont {J.}~\bibnamefont
  {Becker~Tjus}}\ and\ \bibinfo {author} {\bibfnamefont {L.}~\bibnamefont
  {Merten}},\ }\href {https://doi.org/10.1016/j.physrep.2020.05.002} {\bibfield
   {journal} {\bibinfo  {journal} {Physics Reports}\ }\bibinfo {series}
  {Closing in on the Origin of {{Galactic}} Cosmic Rays Using Multimessenger
  Information},\ \textbf {\bibinfo {volume} {872}},\ \bibinfo {pages} {1}
  (\bibinfo {year} {2020})},\ \Eprint {https://arxiv.org/abs/2002.00964}
  {arxiv:2002.00964 [astro-ph.HE]} \BibitemShut {NoStop}%
\bibitem [{\citenamefont {Batista}\ \emph {et~al.}(2019)\citenamefont
  {Batista}, \citenamefont {Biteau}, \citenamefont {Bustamante}, \citenamefont
  {Dolag}, \citenamefont {Engel}, \citenamefont {Fang}, \citenamefont
  {Kampert}, \citenamefont {Kostunin}, \citenamefont {Mostafa}, \citenamefont
  {Murase}, \citenamefont {Sigl}, \citenamefont {Oikonomou}, \citenamefont
  {Olinto}, \citenamefont {Panasyuk}, \citenamefont {Taylor},\ and\
  \citenamefont {Unger}}]{AlvesBatista:2019tlv}%
  \BibitemOpen
  \bibfield  {author} {\bibinfo {author} {\bibfnamefont {R.~A.}\ \bibnamefont
  {Batista}}, \bibinfo {author} {\bibfnamefont {J.}~\bibnamefont {Biteau}},
  \bibinfo {author} {\bibfnamefont {M.}~\bibnamefont {Bustamante}}, \bibinfo
  {author} {\bibfnamefont {K.}~\bibnamefont {Dolag}}, \bibinfo {author}
  {\bibfnamefont {R.}~\bibnamefont {Engel}}, \bibinfo {author} {\bibfnamefont
  {K.}~\bibnamefont {Fang}}, \bibinfo {author} {\bibfnamefont {K.-H.}\
  \bibnamefont {Kampert}}, \bibinfo {author} {\bibfnamefont {D.}~\bibnamefont
  {Kostunin}}, \bibinfo {author} {\bibfnamefont {M.}~\bibnamefont {Mostafa}},
  \bibinfo {author} {\bibfnamefont {K.}~\bibnamefont {Murase}}, \bibinfo
  {author} {\bibfnamefont {G.}~\bibnamefont {Sigl}}, \bibinfo {author}
  {\bibfnamefont {F.}~\bibnamefont {Oikonomou}}, \bibinfo {author}
  {\bibfnamefont {A.~V.}\ \bibnamefont {Olinto}}, \bibinfo {author}
  {\bibfnamefont {M.~I.}\ \bibnamefont {Panasyuk}}, \bibinfo {author}
  {\bibfnamefont {A.}~\bibnamefont {Taylor}},\ and\ \bibinfo {author}
  {\bibfnamefont {M.}~\bibnamefont {Unger}},\ }\href
  {https://doi.org/10.3389/fspas.2019.00023} {\bibfield  {journal} {\bibinfo
  {journal} {Frontiers in Astronomy and Space Sciences}\ }\textbf {\bibinfo
  {volume} {6}},\ \bibinfo {pages} {23} (\bibinfo {year} {2019})},\ \Eprint
  {https://arxiv.org/abs/1903.06714} {arxiv:1903.06714 [astro-ph.HE]}
  \BibitemShut {NoStop}%
\bibitem [{\citenamefont {Drury}(1983)}]{LOC:Drury_1983}%
  \BibitemOpen
  \bibfield  {author} {\bibinfo {author} {\bibfnamefont {L.~O.}\ \bibnamefont
  {Drury}},\ }\href {https://doi.org/10.1088/0034-4885/46/8/002} {\bibfield
  {journal} {\bibinfo  {journal} {Reports on Progress in Physics}\ }\textbf
  {\bibinfo {volume} {46}},\ \bibinfo {pages} {973} (\bibinfo {year}
  {1983})}\BibitemShut {NoStop}%
\bibitem [{\citenamefont {Strong}\ \emph {et~al.}(2007)\citenamefont {Strong},
  \citenamefont {Moskalenko},\ and\ \citenamefont {Ptuskin}}]{Strong:2007nh}%
  \BibitemOpen
  \bibfield  {author} {\bibinfo {author} {\bibfnamefont {A.~W.}\ \bibnamefont
  {Strong}}, \bibinfo {author} {\bibfnamefont {I.~V.}\ \bibnamefont
  {Moskalenko}},\ and\ \bibinfo {author} {\bibfnamefont {V.~S.}\ \bibnamefont
  {Ptuskin}},\ }\href {https://doi.org/10.1146/annurev.nucl.57.090506.123011}
  {\bibfield  {journal} {\bibinfo  {journal} {Annual Review of Nuclear and
  Particle Science}\ }\textbf {\bibinfo {volume} {57}},\ \bibinfo {pages} {285}
  (\bibinfo {year} {2007})},\ \Eprint {https://arxiv.org/abs/astro-ph/0701517}
  {arxiv:astro-ph/0701517} \BibitemShut {NoStop}%
\bibitem [{\citenamefont {Peters}(1961)}]{Peters:1961mxb}%
  \BibitemOpen
  \bibfield  {author} {\bibinfo {author} {\bibfnamefont {B.}~\bibnamefont
  {Peters}},\ }\href {https://doi.org/10.1007/bf02783106} {\bibfield  {journal}
  {\bibinfo  {journal} {Nuovo Cim.}\ }\textbf {\bibinfo {volume} {22}},\
  \bibinfo {pages} {800} (\bibinfo {year} {1961})}\BibitemShut {NoStop}%
\bibitem [{\citenamefont {Panov}\ \emph {et~al.}(2007)\citenamefont {Panov}
  \emph {et~al.}}]{Panov:2006kf}%
  \BibitemOpen
  \bibfield  {author} {\bibinfo {author} {\bibfnamefont {A.~D.}\ \bibnamefont
  {Panov}} \emph {et~al.},\ }\href {https://doi.org/10.3103/S1062873807040168}
  {\bibfield  {journal} {\bibinfo  {journal} {Bull. Russ. Acad. Sci. Phys.}\
  }\textbf {\bibinfo {volume} {71}},\ \bibinfo {pages} {494} (\bibinfo {year}
  {2007})},\ \Eprint {https://arxiv.org/abs/astro-ph/0612377}
  {arXiv:astro-ph/0612377} \BibitemShut {NoStop}%
\bibitem [{\citenamefont {Panov}\ \emph {et~al.}(2009)\citenamefont {Panov},
  \citenamefont {Adams}, \citenamefont {Ahn}, \citenamefont {Bashinzhagyan},
  \citenamefont {Watts}, \citenamefont {Wefel}, \citenamefont {Wu},
  \citenamefont {Ganel}, \citenamefont {Guzik}, \citenamefont {Zatsepin},
  \citenamefont {Isbert}, \citenamefont {Kim}, \citenamefont {Christl},
  \citenamefont {Kouznetsov}, \citenamefont {Panasyuk}, \citenamefont {Seo},
  \citenamefont {Sokolskaya}, \citenamefont {Chang}, \citenamefont {Schmidt},\
  and\ \citenamefont {Fazely}}]{Panov:2009iih}%
  \BibitemOpen
  \bibfield  {author} {\bibinfo {author} {\bibfnamefont {A.~D.}\ \bibnamefont
  {Panov}}, \bibinfo {author} {\bibfnamefont {J.~H.}\ \bibnamefont {Adams}},
  \bibinfo {author} {\bibfnamefont {H.~S.}\ \bibnamefont {Ahn}}, \bibinfo
  {author} {\bibfnamefont {G.~L.}\ \bibnamefont {Bashinzhagyan}}, \bibinfo
  {author} {\bibfnamefont {J.~W.}\ \bibnamefont {Watts}}, \bibinfo {author}
  {\bibfnamefont {J.~P.}\ \bibnamefont {Wefel}}, \bibinfo {author}
  {\bibfnamefont {J.}~\bibnamefont {Wu}}, \bibinfo {author} {\bibfnamefont
  {O.}~\bibnamefont {Ganel}}, \bibinfo {author} {\bibfnamefont {T.~G.}\
  \bibnamefont {Guzik}}, \bibinfo {author} {\bibfnamefont {V.~I.}\ \bibnamefont
  {Zatsepin}}, \bibinfo {author} {\bibfnamefont {I.}~\bibnamefont {Isbert}},
  \bibinfo {author} {\bibfnamefont {K.~C.}\ \bibnamefont {Kim}}, \bibinfo
  {author} {\bibfnamefont {M.}~\bibnamefont {Christl}}, \bibinfo {author}
  {\bibfnamefont {E.~N.}\ \bibnamefont {Kouznetsov}}, \bibinfo {author}
  {\bibfnamefont {M.~I.}\ \bibnamefont {Panasyuk}}, \bibinfo {author}
  {\bibfnamefont {E.~S.}\ \bibnamefont {Seo}}, \bibinfo {author} {\bibfnamefont
  {N.~V.}\ \bibnamefont {Sokolskaya}}, \bibinfo {author} {\bibfnamefont
  {J.}~\bibnamefont {Chang}}, \bibinfo {author} {\bibfnamefont {W.~K.~H.}\
  \bibnamefont {Schmidt}},\ and\ \bibinfo {author} {\bibfnamefont {A.~R.}\
  \bibnamefont {Fazely}},\ }\href {https://doi.org/10.3103/S1062873809050098}
  {\bibfield  {journal} {\bibinfo  {journal} {Bulletin of the Russian Academy
  of Sciences: Physics}\ }\textbf {\bibinfo {volume} {73}},\ \bibinfo {pages}
  {564} (\bibinfo {year} {2009})},\ \Eprint {https://arxiv.org/abs/1101.3246}
  {arxiv:1101.3246 [astro-ph.HE]} \BibitemShut {NoStop}%
\bibitem [{\citenamefont {Ahn}\ \emph {et~al.}(2010)\citenamefont {Ahn} \emph
  {et~al.}}]{Ahn:2010gv}%
  \BibitemOpen
  \bibfield  {author} {\bibinfo {author} {\bibfnamefont {H.~S.}\ \bibnamefont
  {Ahn}} \emph {et~al.},\ }\href {https://doi.org/10.1088/2041-8205/714/1/L89}
  {\bibfield  {journal} {\bibinfo  {journal} {The Astrophysical Journal}\
  }\textbf {\bibinfo {volume} {714}},\ \bibinfo {pages} {L89} (\bibinfo {year}
  {2010})},\ \Eprint {https://arxiv.org/abs/1004.1123} {arxiv:1004.1123
  [astro-ph.HE]} \BibitemShut {NoStop}%
\bibitem [{\citenamefont {Yoon}\ \emph {et~al.}(2017)\citenamefont {Yoon} \emph
  {et~al.}}]{Yoon:2017qjx}%
  \BibitemOpen
  \bibfield  {author} {\bibinfo {author} {\bibfnamefont {Y.~S.}\ \bibnamefont
  {Yoon}} \emph {et~al.},\ }\href {https://doi.org/10.3847/1538-4357/aa68e4}
  {\bibfield  {journal} {\bibinfo  {journal} {The Astrophysical Journal}\
  }\textbf {\bibinfo {volume} {839}},\ \bibinfo {pages} {5} (\bibinfo {year}
  {2017})},\ \Eprint {https://arxiv.org/abs/1704.02512} {arxiv:1704.02512
  [astro-ph.HE]} \BibitemShut {NoStop}%
\bibitem [{\citenamefont {Adriani}\ \emph {et~al.}(2011)\citenamefont {Adriani}
  \emph {et~al.}}]{PAMELA:2011mvy}%
  \BibitemOpen
  \bibfield  {author} {\bibinfo {author} {\bibfnamefont {O.}~\bibnamefont
  {Adriani}} \emph {et~al.} (\bibinfo {collaboration} {PAMELA}),\ }\href
  {https://doi.org/10.1126/science.1199172} {\bibfield  {journal} {\bibinfo
  {journal} {Science}\ }\textbf {\bibinfo {volume} {332}},\ \bibinfo {pages}
  {69} (\bibinfo {year} {2011})},\ \Eprint {https://arxiv.org/abs/1103.4055}
  {arxiv:1103.4055 [astro-ph.HE]} \BibitemShut {NoStop}%
\bibitem [{\citenamefont {Aguilar}\ \emph
  {et~al.}(2015{\natexlab{a}})\citenamefont {Aguilar}, \citenamefont {Aisa},
  \citenamefont {Alpat} \emph {et~al.}}]{AMS:2015azc}%
  \BibitemOpen
  \bibfield  {author} {\bibinfo {author} {\bibfnamefont {M.}~\bibnamefont
  {Aguilar}}, \bibinfo {author} {\bibfnamefont {D.}~\bibnamefont {Aisa}},
  \bibinfo {author} {\bibfnamefont {B.}~\bibnamefont {Alpat}}, \emph {et~al.}
  (\bibinfo {collaboration} {AMS}),\ }\href
  {https://doi.org/10.1103/PhysRevLett.115.211101} {\bibfield  {journal}
  {\bibinfo  {journal} {Physical Review Letters}\ }\textbf {\bibinfo {volume}
  {115}},\ \bibinfo {pages} {211101} (\bibinfo {year}
  {2015}{\natexlab{a}})}\BibitemShut {NoStop}%
\bibitem [{\citenamefont {Aguilar}\ \emph
  {et~al.}(2015{\natexlab{b}})\citenamefont {Aguilar}, \citenamefont {Aisa},
  \citenamefont {Alpat} \emph {et~al.}}]{AMS:2015tnn}%
  \BibitemOpen
  \bibfield  {author} {\bibinfo {author} {\bibfnamefont {M.}~\bibnamefont
  {Aguilar}}, \bibinfo {author} {\bibfnamefont {D.}~\bibnamefont {Aisa}},
  \bibinfo {author} {\bibfnamefont {B.}~\bibnamefont {Alpat}}, \emph {et~al.}
  (\bibinfo {collaboration} {AMS}),\ }\href
  {https://doi.org/10.1103/PhysRevLett.114.171103} {\bibfield  {journal}
  {\bibinfo  {journal} {Physical Review Letters}\ }\textbf {\bibinfo {volume}
  {114}},\ \bibinfo {pages} {171103} (\bibinfo {year}
  {2015}{\natexlab{b}})}\BibitemShut {NoStop}%
\bibitem [{\citenamefont {Atkin}\ \emph {et~al.}(2018)\citenamefont {Atkin}
  \emph {et~al.}}]{Atkin_2018}%
  \BibitemOpen
  \bibfield  {author} {\bibinfo {author} {\bibfnamefont {E.}~\bibnamefont
  {Atkin}} \emph {et~al.},\ }\href {https://doi.org/10.1134/s0021364018130015}
  {\bibfield  {journal} {\bibinfo  {journal} {JETP Letters}\ }\textbf {\bibinfo
  {volume} {108}},\ \bibinfo {pages} {5–12} (\bibinfo {year}
  {2018})}\BibitemShut {NoStop}%
\bibitem [{\citenamefont {An}\ \emph {et~al.}(2019)\citenamefont {An} \emph
  {et~al.}}]{DAMPE:2019gys}%
  \BibitemOpen
  \bibfield  {author} {\bibinfo {author} {\bibfnamefont {Q.}~\bibnamefont {An}}
  \emph {et~al.} (\bibinfo {collaboration} {DAMPE}),\ }\href
  {https://doi.org/10.1126/sciadv.aax3793} {\bibfield  {journal} {\bibinfo
  {journal} {Science Advances}\ }\textbf {\bibinfo {volume} {5}},\ \bibinfo
  {pages} {eaax3793} (\bibinfo {year} {2019})},\ \Eprint
  {https://arxiv.org/abs/1909.12860} {arxiv:1909.12860 [astro-ph.HE]}
  \BibitemShut {NoStop}%
\bibitem [{\citenamefont {Alemanno}\ \emph {et~al.}(2021)\citenamefont
  {Alemanno} \emph {et~al.}}]{Alemanno:2021gpb}%
  \BibitemOpen
  \bibfield  {author} {\bibinfo {author} {\bibfnamefont {F.}~\bibnamefont
  {Alemanno}} \emph {et~al.},\ }\href
  {https://doi.org/10.1103/PhysRevLett.126.201102} {\bibfield  {journal}
  {\bibinfo  {journal} {Phys. Rev. Lett.}\ }\textbf {\bibinfo {volume} {126}},\
  \bibinfo {pages} {201102} (\bibinfo {year} {2021})},\ \Eprint
  {https://arxiv.org/abs/2105.09073} {arxiv:2105.09073 [astro-ph.HE]}
  \BibitemShut {NoStop}%
\bibitem [{\citenamefont
  {Collaboration}(2023)}]{dampecollaboration2023measurement}%
  \BibitemOpen
  \bibfield  {author} {\bibinfo {author} {\bibfnamefont {D.}~\bibnamefont
  {Collaboration}},\ }\href@noop {} {\bibinfo {title} {Measurement of the
  cosmic p+he energy spectrum from 46 gev to 316 tev with the dampe space
  mission}} (\bibinfo {year} {2023}),\ \Eprint
  {https://arxiv.org/abs/2304.00137} {arXiv:2304.00137 [astro-ph.HE]}
  \BibitemShut {NoStop}%
\bibitem [{\citenamefont {Aartsen}\ \emph {et~al.}(2013)\citenamefont {Aartsen}
  \emph {et~al.}}]{IceCube:2013ftu}%
  \BibitemOpen
  \bibfield  {author} {\bibinfo {author} {\bibfnamefont {M.}~\bibnamefont
  {Aartsen}} \emph {et~al.} (\bibinfo {collaboration} {IceCube}),\ }\href
  {https://doi.org/10.1103/PhysRevD.88.042004} {\bibfield  {journal} {\bibinfo
  {journal} {Physical Review D}\ }\textbf {\bibinfo {volume} {88}},\ \bibinfo
  {pages} {042004} (\bibinfo {year} {2013})},\ \Eprint
  {https://arxiv.org/abs/1307.3795} {arxiv:1307.3795 [astro-ph.HE]}
  \BibitemShut {NoStop}%
\bibitem [{\citenamefont {Prosin}\ \emph {et~al.}(2016)\citenamefont {Prosin}
  \emph {et~al.}}]{Prosin:2016rqu}%
  \BibitemOpen
  \bibfield  {author} {\bibinfo {author} {\bibfnamefont {V.~V.}\ \bibnamefont
  {Prosin}} \emph {et~al.},\ }\href
  {https://doi.org/10.1051/epjconf/201612103004} {\bibfield  {journal}
  {\bibinfo  {journal} {EPJ Web Conf.}\ }\textbf {\bibinfo {volume} {121}},\
  \bibinfo {pages} {03004} (\bibinfo {year} {2016})}\BibitemShut {NoStop}%
\bibitem [{\citenamefont {Bergman}\ and\ \citenamefont
  {Belz}(2007)}]{Bergman:2007kn}%
  \BibitemOpen
  \bibfield  {author} {\bibinfo {author} {\bibfnamefont {D.~R.}\ \bibnamefont
  {Bergman}}\ and\ \bibinfo {author} {\bibfnamefont {J.~W.}\ \bibnamefont
  {Belz}},\ }\href {https://doi.org/10.1088/0954-3899/34/10/R01} {\bibfield
  {journal} {\bibinfo  {journal} {Journal of Physics G: Nuclear and Particle
  Physics}\ }\textbf {\bibinfo {volume} {34}},\ \bibinfo {pages} {R359}
  (\bibinfo {year} {2007})},\ \Eprint {https://arxiv.org/abs/0704.3721}
  {arxiv:0704.3721 [astro-ph]} \BibitemShut {NoStop}%
\bibitem [{\citenamefont {Cao}\ \emph {et~al.}(2024)\citenamefont {Cao} \emph
  {et~al.}}]{lhaaso2024allparticle}%
  \BibitemOpen
  \bibfield  {author} {\bibinfo {author} {\bibfnamefont {Z.}~\bibnamefont
  {Cao}} \emph {et~al.} (\bibinfo {collaboration} {LHAASO}),\ }\href@noop {}
  {\bibfield  {journal} {\bibinfo  {journal} {Physical Review Letters}\ }
  (\bibinfo {year} {2024})},\ \Eprint {https://arxiv.org/abs/2403.10010}
  {arXiv:2403.10010 [astro-ph.HE]} \BibitemShut {NoStop}%
\bibitem [{\citenamefont {H{\"o}randel}(2003)}]{Hoerandel:2002yg}%
  \BibitemOpen
  \bibfield  {author} {\bibinfo {author} {\bibfnamefont {J.~R.}\ \bibnamefont
  {H{\"o}randel}},\ }\href {https://doi.org/10.1016/S0927-6505(02)00198-6}
  {\bibfield  {journal} {\bibinfo  {journal} {Astroparticle Physics}\ }\textbf
  {\bibinfo {volume} {19}},\ \bibinfo {pages} {193} (\bibinfo {year} {2003})},\
  \Eprint {https://arxiv.org/abs/astro-ph/0210453} {arxiv:astro-ph/0210453}
  \BibitemShut {NoStop}%
\bibitem [{\citenamefont {Zatsepin}\ and\ \citenamefont
  {Sokolskaya}(2006)}]{Zatsepin:2006ci}%
  \BibitemOpen
  \bibfield  {author} {\bibinfo {author} {\bibfnamefont {V.~I.}\ \bibnamefont
  {Zatsepin}}\ and\ \bibinfo {author} {\bibfnamefont {N.~V.}\ \bibnamefont
  {Sokolskaya}},\ }\href {https://doi.org/10.1051/0004-6361:20065108}
  {\bibfield  {journal} {\bibinfo  {journal} {Astronomy \& Astrophysics}\
  }\textbf {\bibinfo {volume} {458}},\ \bibinfo {pages} {1} (\bibinfo {year}
  {2006})},\ \Eprint {https://arxiv.org/abs/astro-ph/0601475}
  {arxiv:astro-ph/0601475} \BibitemShut {NoStop}%
\bibitem [{\citenamefont {Hillas}(2006)}]{Hillas:2006ms}%
  \BibitemOpen
  \bibfield  {author} {\bibinfo {author} {\bibfnamefont {A.~M.}\ \bibnamefont
  {Hillas}},\ }\bibfield  {journal} {\bibinfo  {journal} {arXiv e-prints}\
  }\href {https://doi.org/10.48550/arXiv.astro-ph/0607109}
  {10.48550/arXiv.astro-ph/0607109} (\bibinfo {year} {2006}),\ \Eprint
  {https://arxiv.org/abs/astro-ph/0607109} {arxiv:astro-ph/0607109}
  \BibitemShut {NoStop}%
\bibitem [{\citenamefont {Gaisser}\ \emph {et~al.}(2013)\citenamefont
  {Gaisser}, \citenamefont {Stanev},\ and\ \citenamefont
  {Tilav}}]{Gaisser:2013bla}%
  \BibitemOpen
  \bibfield  {author} {\bibinfo {author} {\bibfnamefont {T.~K.}\ \bibnamefont
  {Gaisser}}, \bibinfo {author} {\bibfnamefont {T.}~\bibnamefont {Stanev}},\
  and\ \bibinfo {author} {\bibfnamefont {S.}~\bibnamefont {Tilav}},\ }\href
  {https://doi.org/10.1007/s11467-013-0319-7} {\bibfield  {journal} {\bibinfo
  {journal} {Frontiers of Physics}\ }\textbf {\bibinfo {volume} {8}},\ \bibinfo
  {pages} {748} (\bibinfo {year} {2013})},\ \Eprint
  {https://arxiv.org/abs/1303.3565} {arxiv:1303.3565 [astro-ph.HE]}
  \BibitemShut {NoStop}%
\bibitem [{\citenamefont {Stanev}\ \emph {et~al.}(2014)\citenamefont {Stanev},
  \citenamefont {Gaisser},\ and\ \citenamefont {Tilav}}]{Stanev:2014mla}%
  \BibitemOpen
  \bibfield  {author} {\bibinfo {author} {\bibfnamefont {T.}~\bibnamefont
  {Stanev}}, \bibinfo {author} {\bibfnamefont {T.~K.}\ \bibnamefont
  {Gaisser}},\ and\ \bibinfo {author} {\bibfnamefont {S.}~\bibnamefont
  {Tilav}},\ }\href {https://doi.org/10.1016/j.nima.2013.11.094} {\bibfield
  {journal} {\bibinfo  {journal} {Nucl. Instrum. Meth. A}\ }\textbf {\bibinfo
  {volume} {742}},\ \bibinfo {pages} {42} (\bibinfo {year} {2014})}\BibitemShut
  {NoStop}%
\bibitem [{\citenamefont {Thoudam}\ \emph {et~al.}(2016)\citenamefont
  {Thoudam}, \citenamefont {Rachen}, \citenamefont {{van Vliet}}, \citenamefont
  {Achterberg}, \citenamefont {Buitink}, \citenamefont {Falcke},\ and\
  \citenamefont {H{\"o}randel}}]{Thoudam:2016syr}%
  \BibitemOpen
  \bibfield  {author} {\bibinfo {author} {\bibfnamefont {S.}~\bibnamefont
  {Thoudam}}, \bibinfo {author} {\bibfnamefont {J.~P.}\ \bibnamefont {Rachen}},
  \bibinfo {author} {\bibfnamefont {A.}~\bibnamefont {{van Vliet}}}, \bibinfo
  {author} {\bibfnamefont {A.}~\bibnamefont {Achterberg}}, \bibinfo {author}
  {\bibfnamefont {S.}~\bibnamefont {Buitink}}, \bibinfo {author} {\bibfnamefont
  {H.}~\bibnamefont {Falcke}},\ and\ \bibinfo {author} {\bibfnamefont {J.~R.}\
  \bibnamefont {H{\"o}randel}},\ }\href
  {https://doi.org/10.1051/0004-6361/201628894} {\bibfield  {journal} {\bibinfo
   {journal} {Astronomy \& Astrophysics}\ }\textbf {\bibinfo {volume} {595}},\
  \bibinfo {pages} {A33} (\bibinfo {year} {2016})},\ \Eprint
  {https://arxiv.org/abs/1605.03111} {arxiv:1605.03111 [astro-ph.HE]}
  \BibitemShut {NoStop}%
\bibitem [{\citenamefont {Dembinski}\ \emph {et~al.}(2018)\citenamefont
  {Dembinski}, \citenamefont {Engel}, \citenamefont {Fedynitch}, \citenamefont
  {Gaisser}, \citenamefont {Riehn},\ and\ \citenamefont
  {Stanev}}]{Dembinski:2017zsh}%
  \BibitemOpen
  \bibfield  {author} {\bibinfo {author} {\bibfnamefont {H.~P.}\ \bibnamefont
  {Dembinski}}, \bibinfo {author} {\bibfnamefont {R.}~\bibnamefont {Engel}},
  \bibinfo {author} {\bibfnamefont {A.}~\bibnamefont {Fedynitch}}, \bibinfo
  {author} {\bibfnamefont {T.}~\bibnamefont {Gaisser}}, \bibinfo {author}
  {\bibfnamefont {F.}~\bibnamefont {Riehn}},\ and\ \bibinfo {author}
  {\bibfnamefont {T.}~\bibnamefont {Stanev}},\ }\href
  {https://doi.org/10.22323/1.301.0533} {\bibfield  {journal} {\bibinfo
  {journal} {PoS}\ }\textbf {\bibinfo {volume} {ICRC2017}},\ \bibinfo {pages}
  {533} (\bibinfo {year} {2018})},\ \Eprint {https://arxiv.org/abs/1711.11432}
  {arXiv:1711.11432 [astro-ph.HE]} \BibitemShut {NoStop}%
\bibitem [{\citenamefont {Guo}\ and\ \citenamefont {Yuan}(2018)}]{Guo:2017tle}%
  \BibitemOpen
  \bibfield  {author} {\bibinfo {author} {\bibfnamefont {Y.-Q.}\ \bibnamefont
  {Guo}}\ and\ \bibinfo {author} {\bibfnamefont {Q.}~\bibnamefont {Yuan}},\
  }\href {https://doi.org/10.1088/1674-1137/42/7/075103} {\bibfield  {journal}
  {\bibinfo  {journal} {Chinese Physics C}\ }\textbf {\bibinfo {volume} {42}},\
  \bibinfo {pages} {075103} (\bibinfo {year} {2018})},\ \Eprint
  {https://arxiv.org/abs/1701.07136} {arxiv:1701.07136 [astro-ph.HE]}
  \BibitemShut {NoStop}%
\bibitem [{\citenamefont {Yue}\ \emph {et~al.}(2019)\citenamefont {Yue} \emph
  {et~al.}}]{Yue:2019sxt}%
  \BibitemOpen
  \bibfield  {author} {\bibinfo {author} {\bibnamefont {Yue}} \emph {et~al.},\
  }\href {https://doi.org/10.1007/s11467-019-0946-8} {\bibfield  {journal}
  {\bibinfo  {journal} {Frontiers of Physics}\ }\textbf {\bibinfo {volume}
  {15}},\ \bibinfo {pages} {24601} (\bibinfo {year} {2019})},\ \Eprint
  {https://arxiv.org/abs/1909.12857} {arxiv:1909.12857 [astro-ph.HE]}
  \BibitemShut {NoStop}%
\bibitem [{\citenamefont {Potgieter}(2013)}]{Potgieter:2013pdj}%
  \BibitemOpen
  \bibfield  {author} {\bibinfo {author} {\bibfnamefont {M.~S.}\ \bibnamefont
  {Potgieter}},\ }\href {https://doi.org/10.12942/lrsp-2013-3} {\bibfield
  {journal} {\bibinfo  {journal} {Living Reviews in Solar Physics}\ }\textbf
  {\bibinfo {volume} {10}},\ \bibinfo {pages} {3} (\bibinfo {year} {2013})},\
  \Eprint {https://arxiv.org/abs/1306.4421} {arxiv:1306.4421
  [physics.space-ph]} \BibitemShut {NoStop}%
\bibitem [{\citenamefont {Aguilar}\ \emph {et~al.}(2021)\citenamefont {Aguilar}
  \emph {et~al.}}]{AMS:2021nhj}%
  \BibitemOpen
  \bibfield  {author} {\bibinfo {author} {\bibfnamefont {M.}~\bibnamefont
  {Aguilar}} \emph {et~al.} (\bibinfo {collaboration} {AMS}),\ }\href
  {https://doi.org/10.1016/j.physrep.2020.09.003} {\bibfield  {journal}
  {\bibinfo  {journal} {Physics Reports}\ }\bibinfo {series} {The {{Alpha
  Magnetic Spectrometer}} ({{AMS}}) on the {{International Space Station}}:
  {{Part II}} - {{Results}} from the {{First Seven Years}}},\ \textbf {\bibinfo
  {volume} {894}},\ \bibinfo {pages} {1} (\bibinfo {year} {2021})}\BibitemShut
  {NoStop}%
\bibitem [{\citenamefont {Adriani}\ \emph {et~al.}(2022)\citenamefont {Adriani}
  \emph {et~al.}}]{CALET:2022vro}%
  \BibitemOpen
  \bibfield  {author} {\bibinfo {author} {\bibfnamefont {O.}~\bibnamefont
  {Adriani}} \emph {et~al.} (\bibinfo {collaboration} {CALET}),\ }\href
  {https://doi.org/10.1103/PhysRevLett.129.101102} {\bibfield  {journal}
  {\bibinfo  {journal} {Phys. Rev. Lett.}\ }\textbf {\bibinfo {volume} {129}},\
  \bibinfo {pages} {101102} (\bibinfo {year} {2022})},\ \Eprint
  {https://arxiv.org/abs/2209.01302} {arXiv:2209.01302 [astro-ph.HE]}
  \BibitemShut {NoStop}%
\bibitem [{\citenamefont {Adriani}\ \emph {et~al.}(2016)\citenamefont {Adriani}
  \emph {et~al.}}]{PAMELA:2015kyy}%
  \BibitemOpen
  \bibfield  {author} {\bibinfo {author} {\bibfnamefont {O.}~\bibnamefont
  {Adriani}} \emph {et~al.} (\bibinfo {collaboration} {PAMELA}),\ }\href
  {https://doi.org/10.3847/0004-637X/818/1/68} {\bibfield  {journal} {\bibinfo
  {journal} {Astrophys. J.}\ }\textbf {\bibinfo {volume} {818}},\ \bibinfo
  {pages} {68} (\bibinfo {year} {2016})},\ \Eprint
  {https://arxiv.org/abs/1512.06535} {arXiv:1512.06535 [astro-ph.HE]}
  \BibitemShut {NoStop}%
\bibitem [{\citenamefont {Aguilar}\ \emph {et~al.}(2019)\citenamefont
  {Aguilar}, \citenamefont {Ali~Cavasonza}, \citenamefont {Ambrosi} \emph
  {et~al.}}]{AMS:2019nij}%
  \BibitemOpen
  \bibfield  {author} {\bibinfo {author} {\bibfnamefont {M.}~\bibnamefont
  {Aguilar}}, \bibinfo {author} {\bibfnamefont {L.}~\bibnamefont
  {Ali~Cavasonza}}, \bibinfo {author} {\bibfnamefont {G.}~\bibnamefont
  {Ambrosi}}, \emph {et~al.} (\bibinfo {collaboration} {AMS}),\ }\href
  {https://doi.org/10.1103/PhysRevLett.123.181102} {\bibfield  {journal}
  {\bibinfo  {journal} {Physical Review Letters}\ }\textbf {\bibinfo {volume}
  {123}},\ \bibinfo {pages} {181102} (\bibinfo {year} {2019})}\BibitemShut
  {NoStop}%
\bibitem [{\citenamefont {Grebenyuk}\ \emph {et~al.}(2019)\citenamefont
  {Grebenyuk}, \citenamefont {Karmanov}, \citenamefont {Kovalev}, \citenamefont
  {Kudryashov}, \citenamefont {Kurganov}, \citenamefont {Panov}, \citenamefont
  {Podorozhny}, \citenamefont {Tkachenko}, \citenamefont {Tkachev},
  \citenamefont {Turundaevskiy}, \citenamefont {Vasiliev},\ and\ \citenamefont
  {Voronin}}]{GREBENYUK20192546}%
  \BibitemOpen
  \bibfield  {author} {\bibinfo {author} {\bibfnamefont {V.}~\bibnamefont
  {Grebenyuk}}, \bibinfo {author} {\bibfnamefont {D.}~\bibnamefont {Karmanov}},
  \bibinfo {author} {\bibfnamefont {I.}~\bibnamefont {Kovalev}}, \bibinfo
  {author} {\bibfnamefont {I.}~\bibnamefont {Kudryashov}}, \bibinfo {author}
  {\bibfnamefont {A.}~\bibnamefont {Kurganov}}, \bibinfo {author}
  {\bibfnamefont {A.}~\bibnamefont {Panov}}, \bibinfo {author} {\bibfnamefont
  {D.}~\bibnamefont {Podorozhny}}, \bibinfo {author} {\bibfnamefont
  {A.}~\bibnamefont {Tkachenko}}, \bibinfo {author} {\bibfnamefont
  {L.}~\bibnamefont {Tkachev}}, \bibinfo {author} {\bibfnamefont
  {A.}~\bibnamefont {Turundaevskiy}}, \bibinfo {author} {\bibfnamefont
  {O.}~\bibnamefont {Vasiliev}},\ and\ \bibinfo {author} {\bibfnamefont
  {A.}~\bibnamefont {Voronin}},\ }\href
  {https://doi.org/https://doi.org/10.1016/j.asr.2019.10.004} {\bibfield
  {journal} {\bibinfo  {journal} {Advances in Space Research}\ }\textbf
  {\bibinfo {volume} {64}},\ \bibinfo {pages} {2546} (\bibinfo {year}
  {2019})},\ \bibinfo {note} {advances in Cosmic-Ray Astrophysics and Related
  Areas}\BibitemShut {NoStop}%
\bibitem [{\citenamefont {Prosin}\ \emph {et~al.}(2014)\citenamefont {Prosin}
  \emph {et~al.}}]{PROSIN201494}%
  \BibitemOpen
  \bibfield  {author} {\bibinfo {author} {\bibfnamefont {V.}~\bibnamefont
  {Prosin}} \emph {et~al.},\ }\href
  {https://doi.org/https://doi.org/10.1016/j.nima.2013.09.018} {\bibfield
  {journal} {\bibinfo  {journal} {Nuclear Instruments and Methods in Physics
  Research Section A: Accelerators, Spectrometers, Detectors and Associated
  Equipment}\ }\textbf {\bibinfo {volume} {756}},\ \bibinfo {pages} {94}
  (\bibinfo {year} {2014})}\BibitemShut {NoStop}%
\bibitem [{\citenamefont {Abreu}\ \emph {et~al.}(2021)\citenamefont {Abreu}
  \emph {et~al.}}]{PierreAuger:2021hun}%
  \BibitemOpen
  \bibfield  {author} {\bibinfo {author} {\bibfnamefont {P.}~\bibnamefont
  {Abreu}} \emph {et~al.} (\bibinfo {collaboration} {Pierre Auger}),\ }\href
  {https://doi.org/10.1140/epjc/s10052-021-09700-w} {\bibfield  {journal}
  {\bibinfo  {journal} {Eur. Phys. J. C}\ }\textbf {\bibinfo {volume} {81}},\
  \bibinfo {pages} {966} (\bibinfo {year} {2021})},\ \Eprint
  {https://arxiv.org/abs/2109.13400} {arXiv:2109.13400 [astro-ph.HE]}
  \BibitemShut {NoStop}%
\bibitem [{\citenamefont {Morales-Soto}\ \emph {et~al.}(2021)\citenamefont
  {Morales-Soto} \emph {et~al.}}]{HAWC:2021ubt}%
  \BibitemOpen
  \bibfield  {author} {\bibinfo {author} {\bibfnamefont {J.~A.}\ \bibnamefont
  {Morales-Soto}} \emph {et~al.} (\bibinfo {collaboration} {HAWC}),\ }\href
  {https://doi.org/10.22323/1.395.0330} {\bibfield  {journal} {\bibinfo
  {journal} {PoS}\ }\textbf {\bibinfo {volume} {ICRC2021}},\ \bibinfo {pages}
  {330} (\bibinfo {year} {2021})},\ \Eprint {https://arxiv.org/abs/2108.04748}
  {arXiv:2108.04748 [astro-ph.HE]} \BibitemShut {NoStop}%
\bibitem [{\citenamefont {Aartsen}\ \emph {et~al.}(2020)\citenamefont {Aartsen}
  \emph {et~al.}}]{IceCube:2020yct}%
  \BibitemOpen
  \bibfield  {author} {\bibinfo {author} {\bibfnamefont {M.~G.}\ \bibnamefont
  {Aartsen}} \emph {et~al.} (\bibinfo {collaboration} {IceCube}),\ }\href
  {https://doi.org/10.1103/PhysRevD.102.122001} {\bibfield  {journal} {\bibinfo
   {journal} {Phys. Rev. D}\ }\textbf {\bibinfo {volume} {102}},\ \bibinfo
  {pages} {122001} (\bibinfo {year} {2020})},\ \Eprint
  {https://arxiv.org/abs/2006.05215} {arXiv:2006.05215 [astro-ph.HE]}
  \BibitemShut {NoStop}%
\bibitem [{\citenamefont {Finger}(2011)}]{Finger:2011bia}%
  \BibitemOpen
  \bibfield  {author} {\bibinfo {author} {\bibfnamefont {M.~R.}\ \bibnamefont
  {Finger}},\ }\emph {\bibinfo {title} {{Reconstruction of energy spectra for
  different mass groups of high-energy cosmic rays}}},\ \href@noop {} {Ph.D.
  thesis},\ \bibinfo  {school} {KIT, Karlsruhe} (\bibinfo {year}
  {2011})\BibitemShut {NoStop}%
\bibitem [{\citenamefont {Apel}\ \emph {et~al.}(2013)\citenamefont {Apel} \emph
  {et~al.}}]{Apel:2013uni}%
  \BibitemOpen
  \bibfield  {author} {\bibinfo {author} {\bibfnamefont {W.~D.}\ \bibnamefont
  {Apel}} \emph {et~al.},\ }\href
  {https://doi.org/10.1016/j.astropartphys.2013.06.004} {\bibfield  {journal}
  {\bibinfo  {journal} {Astropart. Phys.}\ }\textbf {\bibinfo {volume} {47}},\
  \bibinfo {pages} {54} (\bibinfo {year} {2013})},\ \Eprint
  {https://arxiv.org/abs/1306.6283} {arXiv:1306.6283 [astro-ph.HE]}
  \BibitemShut {NoStop}%
\bibitem [{\citenamefont {Ivanov}(2020)}]{Ivanov:2020rqn}%
  \BibitemOpen
  \bibfield  {author} {\bibinfo {author} {\bibfnamefont {D.}~\bibnamefont
  {Ivanov}} (\bibinfo {collaboration} {Telescope Array}),\ }\href
  {https://doi.org/10.22323/1.358.0298} {\bibfield  {journal} {\bibinfo
  {journal} {PoS}\ }\textbf {\bibinfo {volume} {ICRC2019}},\ \bibinfo {pages}
  {298} (\bibinfo {year} {2020})}\BibitemShut {NoStop}%
\bibitem [{\citenamefont {Abbasi}\ \emph
  {et~al.}(2018{\natexlab{a}})\citenamefont {Abbasi} \emph
  {et~al.}}]{TelescopeArray:2018bya}%
  \BibitemOpen
  \bibfield  {author} {\bibinfo {author} {\bibfnamefont {R.~U.}\ \bibnamefont
  {Abbasi}} \emph {et~al.} (\bibinfo {collaboration} {Telescope Array}),\
  }\href {https://doi.org/10.3847/1538-4357/aada05} {\bibfield  {journal}
  {\bibinfo  {journal} {Astrophys. J.}\ }\textbf {\bibinfo {volume} {865}},\
  \bibinfo {pages} {74} (\bibinfo {year} {2018}{\natexlab{a}})},\ \Eprint
  {https://arxiv.org/abs/1803.01288} {arXiv:1803.01288 [astro-ph.HE]}
  \BibitemShut {NoStop}%
\bibitem [{\citenamefont {Amenomori}\ \emph {et~al.}(2008)\citenamefont
  {Amenomori} \emph {et~al.}}]{TIBETIII:2008qon}%
  \BibitemOpen
  \bibfield  {author} {\bibinfo {author} {\bibfnamefont {M.}~\bibnamefont
  {Amenomori}} \emph {et~al.} (\bibinfo {collaboration} {TIBET III}),\ }\href
  {https://doi.org/10.1086/529514} {\bibfield  {journal} {\bibinfo  {journal}
  {Astrophys. J.}\ }\textbf {\bibinfo {volume} {678}},\ \bibinfo {pages} {1165}
  (\bibinfo {year} {2008})},\ \Eprint {https://arxiv.org/abs/0801.1803}
  {arXiv:0801.1803 [hep-ex]} \BibitemShut {NoStop}%
\bibitem [{\citenamefont {Mayotte}\ \emph {et~al.}(2023)\citenamefont {Mayotte}
  \emph {et~al.}}]{PierreAuger:2023bfx}%
  \BibitemOpen
  \bibfield  {author} {\bibinfo {author} {\bibfnamefont {E.~W.}\ \bibnamefont
  {Mayotte}} \emph {et~al.} (\bibinfo {collaboration} {Pierre Auger}),\ }\href
  {https://doi.org/10.22323/1.444.0365} {\bibfield  {journal} {\bibinfo
  {journal} {PoS}\ }\textbf {\bibinfo {volume} {ICRC2023}},\ \bibinfo {pages}
  {365} (\bibinfo {year} {2023})}\BibitemShut {NoStop}%
\bibitem [{\citenamefont {James}\ and\ \citenamefont
  {Roos}(1975)}]{James:1975dr}%
  \BibitemOpen
  \bibfield  {author} {\bibinfo {author} {\bibfnamefont {F.}~\bibnamefont
  {James}}\ and\ \bibinfo {author} {\bibfnamefont {M.}~\bibnamefont {Roos}},\
  }\href {https://doi.org/10.1016/0010-4655(75)90039-9} {\bibfield  {journal}
  {\bibinfo  {journal} {Comput. Phys. Commun.}\ }\textbf {\bibinfo {volume}
  {10}},\ \bibinfo {pages} {343} (\bibinfo {year} {1975})}\BibitemShut
  {NoStop}%
\bibitem [{\citenamefont {Albert}\ \emph {et~al.}(2022)\citenamefont {Albert}
  \emph {et~al.}}]{HAWC:2022zma}%
  \BibitemOpen
  \bibfield  {author} {\bibinfo {author} {\bibfnamefont {A.}~\bibnamefont
  {Albert}} \emph {et~al.} (\bibinfo {collaboration} {HAWC}),\ }\href
  {https://doi.org/10.1103/PhysRevD.105.063021} {\bibfield  {journal} {\bibinfo
   {journal} {Phys. Rev. D}\ }\textbf {\bibinfo {volume} {105}},\ \bibinfo
  {pages} {063021} (\bibinfo {year} {2022})},\ \Eprint
  {https://arxiv.org/abs/2204.06662} {arXiv:2204.06662 [astro-ph.HE]}
  \BibitemShut {NoStop}%
\bibitem [{\citenamefont {Adriani}\ \emph {et~al.}(2023)\citenamefont {Adriani}
  \emph {et~al.}}]{CALET:2023nif}%
  \BibitemOpen
  \bibfield  {author} {\bibinfo {author} {\bibfnamefont {O.}~\bibnamefont
  {Adriani}} \emph {et~al.} (\bibinfo {collaboration} {CALET}),\ }\href
  {https://doi.org/10.1103/PhysRevLett.130.171002} {\bibfield  {journal}
  {\bibinfo  {journal} {Phys. Rev. Lett.}\ }\textbf {\bibinfo {volume} {130}},\
  \bibinfo {pages} {171002} (\bibinfo {year} {2023})},\ \Eprint
  {https://arxiv.org/abs/2304.14699} {arXiv:2304.14699 [astro-ph.HE]}
  \BibitemShut {NoStop}%
\bibitem [{\citenamefont {Adriani}\ \emph {et~al.}(2020)\citenamefont {Adriani}
  \emph {et~al.}}]{Adriani:2020wyg}%
  \BibitemOpen
  \bibfield  {author} {\bibinfo {author} {\bibfnamefont {O.}~\bibnamefont
  {Adriani}} \emph {et~al.},\ }\href
  {https://doi.org/10.1103/PhysRevLett.125.251102} {\bibfield  {journal}
  {\bibinfo  {journal} {Phys. Rev. Lett.}\ }\textbf {\bibinfo {volume} {125}},\
  \bibinfo {pages} {251102} (\bibinfo {year} {2020})},\ \Eprint
  {https://arxiv.org/abs/2012.10319} {arXiv:2012.10319 [astro-ph.HE]}
  \BibitemShut {NoStop}%
\bibitem [{\citenamefont {Adriani}\ \emph {et~al.}(2021)\citenamefont {Adriani}
  \emph {et~al.}}]{CALET:2021fks}%
  \BibitemOpen
  \bibfield  {author} {\bibinfo {author} {\bibfnamefont {O.}~\bibnamefont
  {Adriani}} \emph {et~al.} (\bibinfo {collaboration} {CALET}),\ }\href
  {https://doi.org/10.1103/PhysRevLett.126.241101} {\bibfield  {journal}
  {\bibinfo  {journal} {Phys. Rev. Lett.}\ }\textbf {\bibinfo {volume} {126}},\
  \bibinfo {pages} {241101} (\bibinfo {year} {2021})},\ \Eprint
  {https://arxiv.org/abs/2106.08036} {arXiv:2106.08036 [astro-ph.HE]}
  \BibitemShut {NoStop}%
\bibitem [{\citenamefont {Abreu}\ \emph {et~al.}(2013)\citenamefont {Abreu},
  \citenamefont {Aglietta}, \citenamefont {Ahlers} \emph
  {et~al.}}]{PierreAuger:2013xim}%
  \BibitemOpen
  \bibfield  {author} {\bibinfo {author} {\bibfnamefont {P.}~\bibnamefont
  {Abreu}}, \bibinfo {author} {\bibfnamefont {M.}~\bibnamefont {Aglietta}},
  \bibinfo {author} {\bibfnamefont {M.}~\bibnamefont {Ahlers}}, \emph {et~al.}
  (\bibinfo {collaboration} {Pierre Auger}),\ }\href
  {https://doi.org/10.1088/1475-7516/2013/02/026} {\bibfield  {journal}
  {\bibinfo  {journal} {Journal of Cosmology and Astroparticle Physics}\
  }\textbf {\bibinfo {volume} {02}}\bibfield  {number} {\bibinfo  {number} {
  (02)},\ \bibinfo {pages} {026}},\ }\Eprint {https://arxiv.org/abs/1301.6637}
  {arxiv:1301.6637 [astro-ph.HE]} \BibitemShut {NoStop}%
\bibitem [{\citenamefont {Aartsen}\ \emph {et~al.}(2019)\citenamefont {Aartsen}
  \emph {et~al.}}]{IceCube:2019hmk}%
  \BibitemOpen
  \bibfield  {author} {\bibinfo {author} {\bibfnamefont {M.~G.}\ \bibnamefont
  {Aartsen}} \emph {et~al.} (\bibinfo {collaboration} {IceCube}),\ }\href
  {https://doi.org/10.1103/PhysRevD.100.082002} {\bibfield  {journal} {\bibinfo
   {journal} {Phys. Rev. D}\ }\textbf {\bibinfo {volume} {100}},\ \bibinfo
  {pages} {082002} (\bibinfo {year} {2019})},\ \Eprint
  {https://arxiv.org/abs/1906.04317} {arXiv:1906.04317 [astro-ph.HE]}
  \BibitemShut {NoStop}%
\bibitem [{\citenamefont {Abbasi}\ \emph
  {et~al.}(2018{\natexlab{b}})\citenamefont {Abbasi} \emph
  {et~al.}}]{TelescopeArray:2018xyi}%
  \BibitemOpen
  \bibfield  {author} {\bibinfo {author} {\bibfnamefont {R.~U.}\ \bibnamefont
  {Abbasi}} \emph {et~al.} (\bibinfo {collaboration} {Telescope Array}),\
  }\href {https://doi.org/10.3847/1538-4357/aabad7} {\bibfield  {journal}
  {\bibinfo  {journal} {Astrophys. J.}\ }\textbf {\bibinfo {volume} {858}},\
  \bibinfo {pages} {76} (\bibinfo {year} {2018}{\natexlab{b}})},\ \Eprint
  {https://arxiv.org/abs/1801.09784} {arXiv:1801.09784 [astro-ph.HE]}
  \BibitemShut {NoStop}%
\bibitem [{\citenamefont {Abbasi}\ \emph {et~al.}(2021)\citenamefont {Abbasi}
  \emph {et~al.}}]{TelescopeArray:2020bfv}%
  \BibitemOpen
  \bibfield  {author} {\bibinfo {author} {\bibfnamefont {R.~U.}\ \bibnamefont
  {Abbasi}} \emph {et~al.} (\bibinfo {collaboration} {Telescope Array}),\
  }\href {https://doi.org/10.3847/1538-4357/abdd30} {\bibfield  {journal}
  {\bibinfo  {journal} {Astrophys. J.}\ }\textbf {\bibinfo {volume} {909}},\
  \bibinfo {pages} {178} (\bibinfo {year} {2021})},\ \Eprint
  {https://arxiv.org/abs/2012.10372} {arXiv:2012.10372 [astro-ph.HE]}
  \BibitemShut {NoStop}%
\bibitem [{\citenamefont {Ahlers}\ and\ \citenamefont
  {Mertsch}(2017)}]{Ahlers:2016rox}%
  \BibitemOpen
  \bibfield  {author} {\bibinfo {author} {\bibfnamefont {M.}~\bibnamefont
  {Ahlers}}\ and\ \bibinfo {author} {\bibfnamefont {P.}~\bibnamefont
  {Mertsch}},\ }\href {https://doi.org/10.1016/j.ppnp.2017.01.004} {\bibfield
  {journal} {\bibinfo  {journal} {Prog. Part. Nucl. Phys.}\ }\textbf {\bibinfo
  {volume} {94}},\ \bibinfo {pages} {184} (\bibinfo {year} {2017})},\ \Eprint
  {https://arxiv.org/abs/1612.01873} {arXiv:1612.01873 [astro-ph.HE]}
  \BibitemShut {NoStop}%
\bibitem [{\citenamefont {Schwadron}\ \emph {et~al.}(2014)\citenamefont
  {Schwadron}, \citenamefont {Adams}, \citenamefont {Christian}, \citenamefont
  {Desiati}, \citenamefont {Frisch}, \citenamefont {Funsten}, \citenamefont
  {Jokipii}, \citenamefont {McComas}, \citenamefont {Moebius},\ and\
  \citenamefont {Zank}}]{doi:10.1126/science.1245026}%
  \BibitemOpen
  \bibfield  {author} {\bibinfo {author} {\bibfnamefont {N.~A.}\ \bibnamefont
  {Schwadron}}, \bibinfo {author} {\bibfnamefont {F.~C.}\ \bibnamefont
  {Adams}}, \bibinfo {author} {\bibfnamefont {E.~R.}\ \bibnamefont
  {Christian}}, \bibinfo {author} {\bibfnamefont {P.}~\bibnamefont {Desiati}},
  \bibinfo {author} {\bibfnamefont {P.}~\bibnamefont {Frisch}}, \bibinfo
  {author} {\bibfnamefont {H.~O.}\ \bibnamefont {Funsten}}, \bibinfo {author}
  {\bibfnamefont {J.~R.}\ \bibnamefont {Jokipii}}, \bibinfo {author}
  {\bibfnamefont {D.~J.}\ \bibnamefont {McComas}}, \bibinfo {author}
  {\bibfnamefont {E.}~\bibnamefont {Moebius}},\ and\ \bibinfo {author}
  {\bibfnamefont {G.~P.}\ \bibnamefont {Zank}},\ }\href
  {https://doi.org/10.1126/science.1245026} {\bibfield  {journal} {\bibinfo
  {journal} {Science}\ }\textbf {\bibinfo {volume} {343}},\ \bibinfo {pages}
  {988} (\bibinfo {year} {2014})},\ \Eprint
  {https://arxiv.org/abs/https://www.science.org/doi/pdf/10.1126/science.1245026}
  {https://www.science.org/doi/pdf/10.1126/science.1245026} \BibitemShut
  {NoStop}%
\bibitem [{\citenamefont {Liu}\ \emph {et~al.}(2019)\citenamefont {Liu},
  \citenamefont {Guo},\ and\ \citenamefont {Yuan}}]{Liu:2018fjy}%
  \BibitemOpen
  \bibfield  {author} {\bibinfo {author} {\bibfnamefont {W.}~\bibnamefont
  {Liu}}, \bibinfo {author} {\bibfnamefont {Y.-Q.}\ \bibnamefont {Guo}},\ and\
  \bibinfo {author} {\bibfnamefont {Q.}~\bibnamefont {Yuan}},\ }\href
  {https://doi.org/10.1088/1475-7516/2019/10/010} {\bibfield  {journal}
  {\bibinfo  {journal} {JCAP}\ }\textbf {\bibinfo {volume} {10}},\ \bibinfo
  {pages} {010}},\ \Eprint {https://arxiv.org/abs/1812.09673} {arxiv:1812.09673
  [astro-ph.HE]} \BibitemShut {NoStop}%
\bibitem [{\citenamefont {Fang}\ \emph {et~al.}(2020)\citenamefont {Fang},
  \citenamefont {Bi},\ and\ \citenamefont {Yin}}]{Fang:2020cru}%
  \BibitemOpen
  \bibfield  {author} {\bibinfo {author} {\bibfnamefont {K.}~\bibnamefont
  {Fang}}, \bibinfo {author} {\bibfnamefont {X.-J.}\ \bibnamefont {Bi}},\ and\
  \bibinfo {author} {\bibfnamefont {P.-F.}\ \bibnamefont {Yin}},\ }\href
  {https://doi.org/10.3847/1538-4357/abb8d7} {\bibfield  {journal} {\bibinfo
  {journal} {Astrophys. J.}\ }\textbf {\bibinfo {volume} {903}},\ \bibinfo
  {pages} {69} (\bibinfo {year} {2020})},\ \Eprint
  {https://arxiv.org/abs/2003.13635} {arXiv:2003.13635 [astro-ph.HE]}
  \BibitemShut {NoStop}%
\bibitem [{\citenamefont {Li}\ \emph {et~al.}(2021)\citenamefont {Li},
  \citenamefont {Yuan}, \citenamefont {Liu},\ and\ \citenamefont
  {Guo}}]{Li:2021meq}%
  \BibitemOpen
  \bibfield  {author} {\bibinfo {author} {\bibfnamefont {A.-f.}\ \bibnamefont
  {Li}}, \bibinfo {author} {\bibfnamefont {Q.}~\bibnamefont {Yuan}}, \bibinfo
  {author} {\bibfnamefont {W.}~\bibnamefont {Liu}},\ and\ \bibinfo {author}
  {\bibfnamefont {Y.-q.}\ \bibnamefont {Guo}},\ }\href@noop {} {\bibfield
  {journal} {\bibinfo  {journal} {arXiv e-prints}\ } (\bibinfo {year}
  {2021})},\ \Eprint {https://arxiv.org/abs/2107.00313} {arxiv:2107.00313
  [astro-ph.HE]} \BibitemShut {NoStop}%
\bibitem [{\citenamefont {Qiao}\ \emph {et~al.}(2022)\citenamefont {Qiao},
  \citenamefont {Luo}, \citenamefont {Yuan},\ and\ \citenamefont
  {Guo}}]{Qiao:2022cge}%
  \BibitemOpen
  \bibfield  {author} {\bibinfo {author} {\bibfnamefont {B.-Q.}\ \bibnamefont
  {Qiao}}, \bibinfo {author} {\bibfnamefont {Q.}~\bibnamefont {Luo}}, \bibinfo
  {author} {\bibfnamefont {Q.}~\bibnamefont {Yuan}},\ and\ \bibinfo {author}
  {\bibfnamefont {Y.-Q.}\ \bibnamefont {Guo}},\ }\href
  {https://doi.org/10.3847/1538-4357/aca7fc} {\bibfield  {journal} {\bibinfo
  {journal} {Astrophys. J.}\ }\textbf {\bibinfo {volume} {942}},\ \bibinfo
  {pages} {13} (\bibinfo {year} {2022})},\ \Eprint
  {https://arxiv.org/abs/2201.06234} {arxiv:2201.06234 [astro-ph.HE]}
  \BibitemShut {NoStop}%
\bibitem [{\citenamefont {Zhang}\ \emph {et~al.}(2021)\citenamefont {Zhang},
  \citenamefont {Qiao}, \citenamefont {Liu}, \citenamefont {Cui}, \citenamefont
  {Yuan},\ and\ \citenamefont {Guo}}]{Zhang:2021ipu}%
  \BibitemOpen
  \bibfield  {author} {\bibinfo {author} {\bibfnamefont {P.-p.}\ \bibnamefont
  {Zhang}}, \bibinfo {author} {\bibfnamefont {B.-q.}\ \bibnamefont {Qiao}},
  \bibinfo {author} {\bibfnamefont {W.}~\bibnamefont {Liu}}, \bibinfo {author}
  {\bibfnamefont {S.-w.}\ \bibnamefont {Cui}}, \bibinfo {author} {\bibfnamefont
  {Q.}~\bibnamefont {Yuan}},\ and\ \bibinfo {author} {\bibfnamefont {Y.-q.}\
  \bibnamefont {Guo}},\ }\href {https://doi.org/10.1088/1475-7516/2021/05/012}
  {\bibfield  {journal} {\bibinfo  {journal} {JCAP}\ }\textbf {\bibinfo
  {volume} {05}},\ \bibinfo {pages} {012}},\ \Eprint
  {https://arxiv.org/abs/2101.00189} {arXiv:2101.00189 [astro-ph.HE]}
  \BibitemShut {NoStop}%
\bibitem [{\citenamefont {Zhang}\ \emph {et~al.}(2022)\citenamefont {Zhang},
  \citenamefont {Liu},\ and\ \citenamefont {Zeng}}]{Zhang:2022pzt}%
  \BibitemOpen
  \bibfield  {author} {\bibinfo {author} {\bibfnamefont {Y.}~\bibnamefont
  {Zhang}}, \bibinfo {author} {\bibfnamefont {S.}~\bibnamefont {Liu}},\ and\
  \bibinfo {author} {\bibfnamefont {H.}~\bibnamefont {Zeng}},\ }\href
  {https://doi.org/10.1093/mnras/stac470} {\bibfield  {journal} {\bibinfo
  {journal} {Mon. Not. Roy. Astron. Soc.}\ }\textbf {\bibinfo {volume} {511}},\
  \bibinfo {pages} {6218} (\bibinfo {year} {2022})},\ \Eprint
  {https://arxiv.org/abs/2202.08491} {arXiv:2202.08491 [astro-ph.HE]}
  \BibitemShut {NoStop}%
\bibitem [{\citenamefont {Scrandis}\ \emph {et~al.}(2021)\citenamefont
  {Scrandis}, \citenamefont {Bowman},\ and\ \citenamefont
  {Seo}}]{Scrandis:2021hwu}%
  \BibitemOpen
  \bibfield  {author} {\bibinfo {author} {\bibfnamefont {R.}~\bibnamefont
  {Scrandis}}, \bibinfo {author} {\bibfnamefont {D.~P.}\ \bibnamefont
  {Bowman}},\ and\ \bibinfo {author} {\bibfnamefont {E.-S.}\ \bibnamefont
  {Seo}},\ }\href {https://doi.org/10.22323/1.395.1220} {\bibfield  {journal}
  {\bibinfo  {journal} {PoS}\ }\textbf {\bibinfo {volume} {ICRC2021}},\
  \bibinfo {pages} {1220} (\bibinfo {year} {2021})}\BibitemShut {NoStop}%
\bibitem [{\citenamefont {Hillas}(2004)}]{Hillas:2004nn}%
  \BibitemOpen
  \bibfield  {author} {\bibinfo {author} {\bibfnamefont {A.~M.}\ \bibnamefont
  {Hillas}},\ }\href {https://doi.org/10.1016/j.nuclphysbps.2004.10.004}
  {\bibfield  {journal} {\bibinfo  {journal} {Nucl. Phys. B Proc. Suppl.}\
  }\textbf {\bibinfo {volume} {136}},\ \bibinfo {pages} {139} (\bibinfo {year}
  {2004})}\BibitemShut {NoStop}%
\bibitem [{\citenamefont {Hillas}(2005)}]{Hillas:2005cs}%
  \BibitemOpen
  \bibfield  {author} {\bibinfo {author} {\bibfnamefont {A.~M.}\ \bibnamefont
  {Hillas}},\ }\href {https://doi.org/10.1088/0954-3899/31/5/R02} {\bibfield
  {journal} {\bibinfo  {journal} {Journal of Physics G: Nuclear and Particle
  Physics}\ }\textbf {\bibinfo {volume} {31}},\ \bibinfo {pages} {R95}
  (\bibinfo {year} {2005})}\BibitemShut {NoStop}%
\bibitem [{\citenamefont {Binns}\ \emph {et~al.}(2008)\citenamefont {Binns},
  \citenamefont {Wiedenbeck}, \citenamefont {Arnould}, \citenamefont
  {Cummings}, \citenamefont {{de Nolfo}}, \citenamefont {Goriely},
  \citenamefont {Israel}, \citenamefont {Leske}, \citenamefont {Mewaldt},
  \citenamefont {Stone},\ and\ \citenamefont {{von
  Rosenvinge}}}]{BINNS2008427}%
  \BibitemOpen
  \bibfield  {author} {\bibinfo {author} {\bibfnamefont {W.}~\bibnamefont
  {Binns}}, \bibinfo {author} {\bibfnamefont {M.}~\bibnamefont {Wiedenbeck}},
  \bibinfo {author} {\bibfnamefont {M.}~\bibnamefont {Arnould}}, \bibinfo
  {author} {\bibfnamefont {A.}~\bibnamefont {Cummings}}, \bibinfo {author}
  {\bibfnamefont {G.}~\bibnamefont {{de Nolfo}}}, \bibinfo {author}
  {\bibfnamefont {S.}~\bibnamefont {Goriely}}, \bibinfo {author} {\bibfnamefont
  {M.}~\bibnamefont {Israel}}, \bibinfo {author} {\bibfnamefont
  {R.}~\bibnamefont {Leske}}, \bibinfo {author} {\bibfnamefont
  {R.}~\bibnamefont {Mewaldt}}, \bibinfo {author} {\bibfnamefont
  {E.}~\bibnamefont {Stone}},\ and\ \bibinfo {author} {\bibfnamefont
  {T.}~\bibnamefont {{von Rosenvinge}}},\ }\href
  {https://doi.org/https://doi.org/10.1016/j.newar.2008.05.008} {\bibfield
  {journal} {\bibinfo  {journal} {New Astronomy Reviews}\ }\textbf {\bibinfo
  {volume} {52}},\ \bibinfo {pages} {427} (\bibinfo {year} {2008})},\ \bibinfo
  {note} {astronomy with Radioactivities. VI}\BibitemShut {NoStop}%
\bibitem [{\citenamefont {Aharonian}\ \emph {et~al.}(2019)\citenamefont
  {Aharonian}, \citenamefont {Yang},\ and\ \citenamefont {de~O\~na
  Wilhelmi}}]{Aharonian:2018oau}%
  \BibitemOpen
  \bibfield  {author} {\bibinfo {author} {\bibfnamefont {F.}~\bibnamefont
  {Aharonian}}, \bibinfo {author} {\bibfnamefont {R.}~\bibnamefont {Yang}},\
  and\ \bibinfo {author} {\bibfnamefont {E.}~\bibnamefont {de~O\~na
  Wilhelmi}},\ }\href {https://doi.org/10.1038/s41550-019-0724-0} {\bibfield
  {journal} {\bibinfo  {journal} {Nature Astron.}\ }\textbf {\bibinfo {volume}
  {3}},\ \bibinfo {pages} {561} (\bibinfo {year} {2019})},\ \Eprint
  {https://arxiv.org/abs/1804.02331} {arXiv:1804.02331 [astro-ph.HE]}
  \BibitemShut {NoStop}%
\bibitem [{\citenamefont {Zhang}\ \emph {et~al.}(2017)\citenamefont {Zhang},
  \citenamefont {Liu},\ and\ \citenamefont {Yuan}}]{Zhang:2017ksy}%
  \BibitemOpen
  \bibfield  {author} {\bibinfo {author} {\bibfnamefont {Y.}~\bibnamefont
  {Zhang}}, \bibinfo {author} {\bibfnamefont {S.}~\bibnamefont {Liu}},\ and\
  \bibinfo {author} {\bibfnamefont {Q.}~\bibnamefont {Yuan}},\ }\href
  {https://doi.org/10.3847/2041-8213/aa7de1} {\bibfield  {journal} {\bibinfo
  {journal} {Astrophys. J. Lett.}\ }\textbf {\bibinfo {volume} {844}},\
  \bibinfo {pages} {L3} (\bibinfo {year} {2017})},\ \Eprint
  {https://arxiv.org/abs/1707.00262} {arXiv:1707.00262 [astro-ph.HE]}
  \BibitemShut {NoStop}%
\bibitem [{\citenamefont {Zhang}\ and\ \citenamefont
  {Liu}(2019)}]{Zhang:2018fds}%
  \BibitemOpen
  \bibfield  {author} {\bibinfo {author} {\bibfnamefont {Y.}~\bibnamefont
  {Zhang}}\ and\ \bibinfo {author} {\bibfnamefont {S.}~\bibnamefont {Liu}},\
  }\href {https://doi.org/10.1093/mnras/sty3136} {\bibfield  {journal}
  {\bibinfo  {journal} {Mon. Not. Roy. Astron. Soc.}\ }\textbf {\bibinfo
  {volume} {482}},\ \bibinfo {pages} {5268} (\bibinfo {year} {2019})},\ \Eprint
  {https://arxiv.org/abs/1812.08395} {arXiv:1812.08395 [astro-ph.HE]}
  \BibitemShut {NoStop}%
\bibitem [{\citenamefont {{Jokipii}}\ and\ \citenamefont
  {{Morfill}}(1987)}]{1987ApJ/312/170J}%
  \BibitemOpen
  \bibfield  {author} {\bibinfo {author} {\bibfnamefont {J.~R.}\ \bibnamefont
  {{Jokipii}}}\ and\ \bibinfo {author} {\bibfnamefont {G.}~\bibnamefont
  {{Morfill}}},\ }\href {https://doi.org/10.1086/164857} {\bibfield  {journal}
  {\bibinfo  {journal} {\apj}\ }\textbf {\bibinfo {volume} {312}},\ \bibinfo
  {pages} {170} (\bibinfo {year} {1987})}\BibitemShut {NoStop}%
\bibitem [{\citenamefont {{Zirakashvili}}\ and\ \citenamefont
  {{V{\"o}lk}}(2006)}]{2006AdSpR/37/1923Z}%
  \BibitemOpen
  \bibfield  {author} {\bibinfo {author} {\bibfnamefont {V.~N.}\ \bibnamefont
  {{Zirakashvili}}}\ and\ \bibinfo {author} {\bibfnamefont {H.~J.}\
  \bibnamefont {{V{\"o}lk}}},\ }\href
  {https://doi.org/10.1016/j.asr.2005.06.013} {\bibfield  {journal} {\bibinfo
  {journal} {Advances in Space Research}\ }\textbf {\bibinfo {volume} {37}},\
  \bibinfo {pages} {1923} (\bibinfo {year} {2006})}\BibitemShut {NoStop}%
\bibitem [{\citenamefont {Tomassetti}(2012)}]{Tomassetti:2012ga}%
  \BibitemOpen
  \bibfield  {author} {\bibinfo {author} {\bibfnamefont {N.}~\bibnamefont
  {Tomassetti}},\ }\href {https://doi.org/10.1088/2041-8205/752/1/L13}
  {\bibfield  {journal} {\bibinfo  {journal} {Astrophys. J. Lett.}\ }\textbf
  {\bibinfo {volume} {752}},\ \bibinfo {pages} {L13} (\bibinfo {year}
  {2012})},\ \Eprint {https://arxiv.org/abs/1204.4492} {arXiv:1204.4492
  [astro-ph.HE]} \BibitemShut {NoStop}%
\bibitem [{\citenamefont {Guo}\ \emph {et~al.}(2016)\citenamefont {Guo},
  \citenamefont {Tian},\ and\ \citenamefont {Jin}}]{Guo:2015csa}%
  \BibitemOpen
  \bibfield  {author} {\bibinfo {author} {\bibfnamefont {Y.-Q.}\ \bibnamefont
  {Guo}}, \bibinfo {author} {\bibfnamefont {Z.}~\bibnamefont {Tian}},\ and\
  \bibinfo {author} {\bibfnamefont {C.}~\bibnamefont {Jin}},\ }\href
  {https://doi.org/10.3847/0004-637X/819/1/54} {\bibfield  {journal} {\bibinfo
  {journal} {Astrophys. J.}\ }\textbf {\bibinfo {volume} {819}},\ \bibinfo
  {pages} {54} (\bibinfo {year} {2016})},\ \Eprint
  {https://arxiv.org/abs/1509.08227} {arXiv:1509.08227 [astro-ph.HE]}
  \BibitemShut {NoStop}%
\bibitem [{\citenamefont {Zhao}\ \emph {et~al.}(2021)\citenamefont {Zhao},
  \citenamefont {Fang},\ and\ \citenamefont {Bi}}]{Zhao:2021yzf}%
  \BibitemOpen
  \bibfield  {author} {\bibinfo {author} {\bibfnamefont {M.-J.}\ \bibnamefont
  {Zhao}}, \bibinfo {author} {\bibfnamefont {K.}~\bibnamefont {Fang}},\ and\
  \bibinfo {author} {\bibfnamefont {X.-J.}\ \bibnamefont {Bi}},\ }\href
  {https://doi.org/10.1103/PhysRevD.104.123001} {\bibfield  {journal} {\bibinfo
   {journal} {Phys. Rev. D}\ }\textbf {\bibinfo {volume} {104}},\ \bibinfo
  {pages} {123001} (\bibinfo {year} {2021})},\ \Eprint
  {https://arxiv.org/abs/2109.04112} {arXiv:2109.04112 [astro-ph.HE]}
  \BibitemShut {NoStop}%
\bibitem [{\citenamefont {Arnould}\ \emph {et~al.}(2007)\citenamefont
  {Arnould}, \citenamefont {Goriely},\ and\ \citenamefont
  {Takahashi}}]{Arnould:2007gh}%
  \BibitemOpen
  \bibfield  {author} {\bibinfo {author} {\bibfnamefont {M.}~\bibnamefont
  {Arnould}}, \bibinfo {author} {\bibfnamefont {S.}~\bibnamefont {Goriely}},\
  and\ \bibinfo {author} {\bibfnamefont {K.}~\bibnamefont {Takahashi}},\ }\href
  {https://doi.org/10.1016/j.physrep.2007.06.002} {\bibfield  {journal}
  {\bibinfo  {journal} {Physics Reports}\ }\textbf {\bibinfo {volume} {450}},\
  \bibinfo {pages} {97} (\bibinfo {year} {2007})},\ \Eprint
  {https://arxiv.org/abs/0705.4512} {arxiv:0705.4512 [astro-ph]} \BibitemShut
  {NoStop}%
\bibitem [{\citenamefont {Engelmann}\ \emph {et~al.}(1990)\citenamefont
  {Engelmann}, \citenamefont {Ferrando}, \citenamefont {Soutoul}, \citenamefont
  {Goret},\ and\ \citenamefont {Juliusson}}]{Engelmann:1990zz}%
  \BibitemOpen
  \bibfield  {author} {\bibinfo {author} {\bibfnamefont {J.~J.}\ \bibnamefont
  {Engelmann}}, \bibinfo {author} {\bibfnamefont {P.}~\bibnamefont {Ferrando}},
  \bibinfo {author} {\bibfnamefont {A.}~\bibnamefont {Soutoul}}, \bibinfo
  {author} {\bibfnamefont {P.}~\bibnamefont {Goret}},\ and\ \bibinfo {author}
  {\bibfnamefont {E.}~\bibnamefont {Juliusson}},\ }\href@noop {} {\bibfield
  {journal} {\bibinfo  {journal} {Astron. Astrophys.}\ }\textbf {\bibinfo
  {volume} {233}},\ \bibinfo {pages} {96} (\bibinfo {year} {1990})}\BibitemShut
  {NoStop}%
\bibitem [{\citenamefont {Walsh}(2020)}]{Walsh:2020xon}%
  \BibitemOpen
  \bibfield  {author} {\bibinfo {author} {\bibfnamefont {N.~E.}\ \bibnamefont
  {Walsh}},\ }\emph {\bibinfo {title} {{SuperTIGER Elemental Abundances for the
  Charge Range $41 \leq Z \leq 56$}}},\ \href
  {https://doi.org/10.7936/42be-8069} {Ph.D. thesis},\ \bibinfo  {school}
  {Washington U., St. Louis} (\bibinfo {year} {2020})\BibitemShut {NoStop}%
\bibitem [{\citenamefont {Stanev}\ \emph {et~al.}(2000)\citenamefont {Stanev},
  \citenamefont {Engel}, \citenamefont {Mucke}, \citenamefont {Protheroe},\
  and\ \citenamefont {Rachen}}]{Stanev:2000fb}%
  \BibitemOpen
  \bibfield  {author} {\bibinfo {author} {\bibfnamefont {T.}~\bibnamefont
  {Stanev}}, \bibinfo {author} {\bibfnamefont {R.}~\bibnamefont {Engel}},
  \bibinfo {author} {\bibfnamefont {A.}~\bibnamefont {Mucke}}, \bibinfo
  {author} {\bibfnamefont {R.~J.}\ \bibnamefont {Protheroe}},\ and\ \bibinfo
  {author} {\bibfnamefont {J.~P.}\ \bibnamefont {Rachen}},\ }\href
  {https://doi.org/10.1103/PhysRevD.62.093005} {\bibfield  {journal} {\bibinfo
  {journal} {Phys. Rev. D}\ }\textbf {\bibinfo {volume} {62}},\ \bibinfo
  {pages} {093005} (\bibinfo {year} {2000})},\ \Eprint
  {https://arxiv.org/abs/astro-ph/0003484} {arXiv:astro-ph/0003484}
  \BibitemShut {NoStop}%
\bibitem [{\citenamefont {Lemoine}(2005)}]{Lemoine_2005}%
  \BibitemOpen
  \bibfield  {author} {\bibinfo {author} {\bibfnamefont {M.}~\bibnamefont
  {Lemoine}},\ }\bibfield  {journal} {\bibinfo  {journal} {Physical Review D}\
  }\textbf {\bibinfo {volume} {71}},\ \href
  {https://doi.org/10.1103/physrevd.71.083007} {10.1103/physrevd.71.083007}
  (\bibinfo {year} {2005})\BibitemShut {NoStop}%
\bibitem [{\citenamefont {Berezinsky}\ and\ \citenamefont
  {Gazizov}(2006)}]{Berezinsky_2006}%
  \BibitemOpen
  \bibfield  {author} {\bibinfo {author} {\bibfnamefont {V.}~\bibnamefont
  {Berezinsky}}\ and\ \bibinfo {author} {\bibfnamefont {A.~Z.}\ \bibnamefont
  {Gazizov}},\ }\href {https://doi.org/10.1086/502626} {\bibfield  {journal}
  {\bibinfo  {journal} {The Astrophysical Journal}\ }\textbf {\bibinfo {volume}
  {643}},\ \bibinfo {pages} {8–13} (\bibinfo {year} {2006})}\BibitemShut
  {NoStop}%
\bibitem [{\citenamefont {Mollerach}\ and\ \citenamefont
  {Roulet}(2013)}]{Mollerach_2013}%
  \BibitemOpen
  \bibfield  {author} {\bibinfo {author} {\bibfnamefont {S.}~\bibnamefont
  {Mollerach}}\ and\ \bibinfo {author} {\bibfnamefont {E.}~\bibnamefont
  {Roulet}},\ }\href {https://doi.org/10.1088/1475-7516/2013/10/013} {\bibfield
   {journal} {\bibinfo  {journal} {Journal of Cosmology and Astroparticle
  Physics}\ }\textbf {\bibinfo {volume} {2013}}\bibinfo  {number} { (10)},\
  \bibinfo {pages} {013–013}}\BibitemShut {NoStop}%
\bibitem [{\citenamefont {Mollerach}\ and\ \citenamefont
  {Roulet}(2019)}]{Mollerach:2018lkt}%
  \BibitemOpen
\bibfield  {number} {  }\bibfield  {author} {\bibinfo {author} {\bibfnamefont
  {S.}~\bibnamefont {Mollerach}}\ and\ \bibinfo {author} {\bibfnamefont
  {E.}~\bibnamefont {Roulet}},\ }\href
  {https://doi.org/10.1088/1475-7516/2019/03/017} {\bibfield  {journal}
  {\bibinfo  {journal} {Journal of Cosmology and Astroparticle Physics}\
  }\textbf {\bibinfo {volume} {03}}\bibfield  {number} {\bibinfo  {number} {
  (03)},\ \bibinfo {pages} {017}},\ }\Eprint {https://arxiv.org/abs/1812.04026}
  {arxiv:1812.04026 [astro-ph.HE]} \BibitemShut {NoStop}%
\bibitem [{\citenamefont {Halim}\ \emph {et~al.}(2023)\citenamefont {Halim}
  \emph {et~al.}}]{PierreAuger:2022atd}%
  \BibitemOpen
  \bibfield  {author} {\bibinfo {author} {\bibfnamefont {A.~A.}\ \bibnamefont
  {Halim}} \emph {et~al.} (\bibinfo {collaboration} {Pierre Auger}),\ }\href
  {https://doi.org/10.1088/1475-7516/2023/05/024} {\bibfield  {journal}
  {\bibinfo  {journal} {JCAP}\ }\textbf {\bibinfo {volume} {05}},\ \bibinfo
  {pages} {024}},\ \Eprint {https://arxiv.org/abs/2211.02857} {arxiv:2211.02857
  [astro-ph.HE]} \BibitemShut {NoStop}%
\bibitem [{\citenamefont {Abdul~Halim}\ \emph {et~al.}(2024)\citenamefont
  {Abdul~Halim} \emph {et~al.}}]{PierreAuger:2023htc}%
  \BibitemOpen
  \bibfield  {author} {\bibinfo {author} {\bibfnamefont {A.}~\bibnamefont
  {Abdul~Halim}} \emph {et~al.} (\bibinfo {collaboration} {Pierre Auger}),\
  }\href {https://doi.org/10.1088/1475-7516/2024/01/022} {\bibfield  {journal}
  {\bibinfo  {journal} {JCAP}\ }\textbf {\bibinfo {volume} {01}},\ \bibinfo
  {pages} {022}},\ \Eprint {https://arxiv.org/abs/2305.16693} {arxiv:2305.16693
  [astro-ph.HE]} \BibitemShut {NoStop}%
\bibitem [{\citenamefont {Salamida}(2023)}]{Salamida:2023qpx}%
  \BibitemOpen
  \bibfield  {author} {\bibinfo {author} {\bibfnamefont {F.}~\bibnamefont
  {Salamida}} (\bibinfo {collaboration} {Pierre Auger}),\ }\href@noop {}
  {\bibfield  {journal} {\bibinfo  {journal} {PoS}\ }\textbf {\bibinfo {volume}
  {ICRC2023}},\ \bibinfo {pages} {016} (\bibinfo {year} {2023})},\ \Eprint
  {https://arxiv.org/abs/2312.14673} {arXiv:2312.14673 [astro-ph.HE]}
  \BibitemShut {NoStop}%
\bibitem [{\citenamefont {Aloisio}(2023)}]{Aloisio:2022xzy}%
  \BibitemOpen
  \bibfield  {author} {\bibinfo {author} {\bibfnamefont {R.}~\bibnamefont
  {Aloisio}},\ }\href {https://doi.org/10.1088/1742-6596/2429/1/012008}
  {\bibfield  {journal} {\bibinfo  {journal} {J. Phys. Conf. Ser.}\ }\textbf
  {\bibinfo {volume} {2429}},\ \bibinfo {pages} {012008} (\bibinfo {year}
  {2023})},\ \Eprint {https://arxiv.org/abs/2212.01600} {arXiv:2212.01600
  [astro-ph.HE]} \BibitemShut {NoStop}%
\bibitem [{\citenamefont {Coleman}\ \emph {et~al.}(2023)\citenamefont {Coleman}
  \emph {et~al.}}]{Coleman:2022abf}%
  \BibitemOpen
  \bibfield  {author} {\bibinfo {author} {\bibfnamefont {A.}~\bibnamefont
  {Coleman}} \emph {et~al.},\ }\href
  {https://doi.org/10.1016/j.astropartphys.2023.102819} {\bibfield  {journal}
  {\bibinfo  {journal} {Astropart. Phys.}\ }\textbf {\bibinfo {volume} {149}},\
  \bibinfo {pages} {102819} (\bibinfo {year} {2023})},\ \Eprint
  {https://arxiv.org/abs/2205.05845} {arXiv:2205.05845 [astro-ph.HE]}
  \BibitemShut {NoStop}%
\bibitem [{\citenamefont {Berezinsky}(2007)}]{Berezinsky:2007zza}%
  \BibitemOpen
  \bibfield  {author} {\bibinfo {author} {\bibfnamefont {V.}~\bibnamefont
  {Berezinsky}},\ }\href {https://doi.org/10.1007/s10509-007-9402-2} {\bibfield
   {journal} {\bibinfo  {journal} {Astrophysics and Space Science}\ }\textbf
  {\bibinfo {volume} {309}},\ \bibinfo {pages} {453} (\bibinfo {year}
  {2007})}\BibitemShut {NoStop}%
\bibitem [{\citenamefont {Aloisio}\ \emph {et~al.}(2014)\citenamefont
  {Aloisio}, \citenamefont {Berezinsky},\ and\ \citenamefont
  {Blasi}}]{Aloisio:2013hya}%
  \BibitemOpen
  \bibfield  {author} {\bibinfo {author} {\bibfnamefont {R.}~\bibnamefont
  {Aloisio}}, \bibinfo {author} {\bibfnamefont {V.}~\bibnamefont
  {Berezinsky}},\ and\ \bibinfo {author} {\bibfnamefont {P.}~\bibnamefont
  {Blasi}},\ }\href {https://doi.org/10.1088/1475-7516/2014/10/020} {\bibfield
  {journal} {\bibinfo  {journal} {JCAP}\ }\textbf {\bibinfo {volume} {10}},\
  \bibinfo {pages} {020}},\ \Eprint {https://arxiv.org/abs/1312.7459}
  {arXiv:1312.7459 [astro-ph.HE]} \BibitemShut {NoStop}%
\bibitem [{\citenamefont {Schoo}\ \emph {et~al.}(2016)\citenamefont {Schoo}
  \emph {et~al.}}]{KASCADE-Grande:2015svq}%
  \BibitemOpen
  \bibfield  {author} {\bibinfo {author} {\bibfnamefont {S.}~\bibnamefont
  {Schoo}} \emph {et~al.} (\bibinfo {collaboration} {KASCADE-Grande}),\ }\href
  {https://doi.org/10.22323/1.236.0263} {\bibfield  {journal} {\bibinfo
  {journal} {PoS}\ }\textbf {\bibinfo {volume} {ICRC2015}},\ \bibinfo {pages}
  {263} (\bibinfo {year} {2016})}\BibitemShut {NoStop}%
\bibitem [{\citenamefont {Berezhko}\ and\ \citenamefont
  {Voelk}(2007)}]{Berezhko:2007gh}%
  \BibitemOpen
  \bibfield  {author} {\bibinfo {author} {\bibfnamefont {E.~G.}\ \bibnamefont
  {Berezhko}}\ and\ \bibinfo {author} {\bibfnamefont {H.~J.}\ \bibnamefont
  {Voelk}},\ }\href {https://doi.org/10.1086/518737} {\bibfield  {journal}
  {\bibinfo  {journal} {Astrophys. J. Lett.}\ }\textbf {\bibinfo {volume}
  {661}},\ \bibinfo {pages} {L175} (\bibinfo {year} {2007})},\ \Eprint
  {https://arxiv.org/abs/0704.1715} {arXiv:0704.1715 [astro-ph]} \BibitemShut
  {NoStop}%
\bibitem [{\citenamefont {Sveshnikova}\ \emph {et~al.}(2015)\citenamefont
  {Sveshnikova}, \citenamefont {Kuzmichev}, \citenamefont {Korosteleva},
  \citenamefont {Prosin},\ and\ \citenamefont {Ptuskin}}]{Sveshnikova:2014yes}%
  \BibitemOpen
  \bibfield  {author} {\bibinfo {author} {\bibfnamefont {L.}~\bibnamefont
  {Sveshnikova}}, \bibinfo {author} {\bibfnamefont {L.}~\bibnamefont
  {Kuzmichev}}, \bibinfo {author} {\bibfnamefont {E.}~\bibnamefont
  {Korosteleva}}, \bibinfo {author} {\bibfnamefont {V.}~\bibnamefont
  {Prosin}},\ and\ \bibinfo {author} {\bibfnamefont {V.~S.}\ \bibnamefont
  {Ptuskin}},\ }\href {https://doi.org/10.1016/j.nuclphysbps.2014.10.025}
  {\bibfield  {journal} {\bibinfo  {journal} {Nuclear Physics B - Proceedings
  Supplements}\ }\bibinfo {series} {Cosmic {{Ray Origin}} {\textendash}
  {{Beyond}} the {{Standard Models}}},\ \textbf {\bibinfo {volume}
  {256--257}},\ \bibinfo {pages} {218} (\bibinfo {year} {2015})}\BibitemShut
  {NoStop}%
\bibitem [{\citenamefont {Giacinti}\ \emph {et~al.}(2015)\citenamefont
  {Giacinti}, \citenamefont {Kachelrie\ss{}},\ and\ \citenamefont
  {Semikoz}}]{Giacinti:2015hva}%
  \BibitemOpen
  \bibfield  {author} {\bibinfo {author} {\bibfnamefont {G.}~\bibnamefont
  {Giacinti}}, \bibinfo {author} {\bibfnamefont {M.}~\bibnamefont
  {Kachelrie\ss{}}},\ and\ \bibinfo {author} {\bibfnamefont {D.~V.}\
  \bibnamefont {Semikoz}},\ }\href {https://doi.org/10.1103/PhysRevD.91.083009}
  {\bibfield  {journal} {\bibinfo  {journal} {Phys. Rev. D}\ }\textbf {\bibinfo
  {volume} {91}},\ \bibinfo {pages} {083009} (\bibinfo {year} {2015})},\
  \Eprint {https://arxiv.org/abs/1502.01608} {arXiv:1502.01608 [astro-ph.HE]}
  \BibitemShut {NoStop}%
\bibitem [{\citenamefont {Rachen}\ \emph {et~al.}(1993)\citenamefont {Rachen},
  \citenamefont {Stanev},\ and\ \citenamefont {Biermann}}]{Rachen:1993gf}%
  \BibitemOpen
  \bibfield  {author} {\bibinfo {author} {\bibfnamefont {J.~P.}\ \bibnamefont
  {Rachen}}, \bibinfo {author} {\bibfnamefont {T.}~\bibnamefont {Stanev}},\
  and\ \bibinfo {author} {\bibfnamefont {P.~L.}\ \bibnamefont {Biermann}},\
  }\href@noop {} {\bibfield  {journal} {\bibinfo  {journal} {Astron.
  Astrophys.}\ }\textbf {\bibinfo {volume} {273}},\ \bibinfo {pages} {377}
  (\bibinfo {year} {1993})},\ \Eprint {https://arxiv.org/abs/astro-ph/9302005}
  {arXiv:astro-ph/9302005} \BibitemShut {NoStop}%
\bibitem [{\citenamefont {Wibig}\ and\ \citenamefont
  {Wolfendale}(2005)}]{Wibig:2004ye}%
  \BibitemOpen
  \bibfield  {author} {\bibinfo {author} {\bibfnamefont {T.}~\bibnamefont
  {Wibig}}\ and\ \bibinfo {author} {\bibfnamefont {A.~W.}\ \bibnamefont
  {Wolfendale}},\ }\href {https://doi.org/10.1088/0954-3899/31/3/005}
  {\bibfield  {journal} {\bibinfo  {journal} {J. Phys. G}\ }\textbf {\bibinfo
  {volume} {31}},\ \bibinfo {pages} {255} (\bibinfo {year} {2005})},\ \Eprint
  {https://arxiv.org/abs/astro-ph/0410624} {arXiv:astro-ph/0410624}
  \BibitemShut {NoStop}%
\bibitem [{\citenamefont {Aab}\ \emph {et~al.}(2020)\citenamefont {Aab} \emph
  {et~al.}}]{PierreAuger:2020fbi}%
  \BibitemOpen
  \bibfield  {author} {\bibinfo {author} {\bibfnamefont {A.}~\bibnamefont
  {Aab}} \emph {et~al.} (\bibinfo {collaboration} {Pierre Auger}),\ }\href
  {https://doi.org/10.3847/1538-4357/ab7236} {\bibfield  {journal} {\bibinfo
  {journal} {Astrophys. J.}\ }\textbf {\bibinfo {volume} {891}},\ \bibinfo
  {pages} {142} (\bibinfo {year} {2020})},\ \Eprint
  {https://arxiv.org/abs/2002.06172} {arXiv:2002.06172 [astro-ph.HE]}
  \BibitemShut {NoStop}%
\bibitem [{\citenamefont {Giacinti}\ \emph {et~al.}(2012)\citenamefont
  {Giacinti}, \citenamefont {Kachelriess}, \citenamefont {Semikoz},\ and\
  \citenamefont {Sigl}}]{Giacinti:2011ww}%
  \BibitemOpen
  \bibfield  {author} {\bibinfo {author} {\bibfnamefont {G.}~\bibnamefont
  {Giacinti}}, \bibinfo {author} {\bibfnamefont {M.}~\bibnamefont
  {Kachelriess}}, \bibinfo {author} {\bibfnamefont {D.~V.}\ \bibnamefont
  {Semikoz}},\ and\ \bibinfo {author} {\bibfnamefont {G.}~\bibnamefont
  {Sigl}},\ }\href {https://doi.org/10.1088/1475-7516/2012/07/031} {\bibfield
  {journal} {\bibinfo  {journal} {JCAP}\ }\textbf {\bibinfo {volume} {07}},\
  \bibinfo {pages} {031}},\ \Eprint {https://arxiv.org/abs/1112.5599}
  {arXiv:1112.5599 [astro-ph.HE]} \BibitemShut {NoStop}%
\bibitem [{\citenamefont {Abreu}\ \emph {et~al.}(2012)\citenamefont {Abreu}
  \emph {et~al.}}]{PierreAuger:2012gro}%
  \BibitemOpen
  \bibfield  {author} {\bibinfo {author} {\bibfnamefont {P.}~\bibnamefont
  {Abreu}} \emph {et~al.} (\bibinfo {collaboration} {Pierre Auger}),\ }\href
  {https://doi.org/10.1088/2041-8205/762/1/L13} {\bibfield  {journal} {\bibinfo
   {journal} {Astrophys. J. Lett.}\ }\textbf {\bibinfo {volume} {762}},\
  \bibinfo {pages} {L13} (\bibinfo {year} {2012})},\ \Eprint
  {https://arxiv.org/abs/1212.3083} {arXiv:1212.3083 [astro-ph.HE]}
  \BibitemShut {NoStop}%
\bibitem [{\citenamefont {Aab}\ \emph {et~al.}(2014{\natexlab{a}})\citenamefont
  {Aab} \emph {et~al.}}]{PierreAuger:2014sui}%
  \BibitemOpen
  \bibfield  {author} {\bibinfo {author} {\bibfnamefont {A.}~\bibnamefont
  {Aab}} \emph {et~al.} (\bibinfo {collaboration} {Pierre Auger}),\ }\href
  {https://doi.org/10.1103/PhysRevD.90.122005} {\bibfield  {journal} {\bibinfo
  {journal} {Phys. Rev. D}\ }\textbf {\bibinfo {volume} {90}},\ \bibinfo
  {pages} {122005} (\bibinfo {year} {2014}{\natexlab{a}})},\ \Eprint
  {https://arxiv.org/abs/1409.4809} {arXiv:1409.4809 [astro-ph.HE]}
  \BibitemShut {NoStop}%
\bibitem [{\citenamefont {Aab}\ \emph {et~al.}(2014{\natexlab{b}})\citenamefont
  {Aab} \emph {et~al.}}]{PierreAuger:2014gko}%
  \BibitemOpen
  \bibfield  {author} {\bibinfo {author} {\bibfnamefont {A.}~\bibnamefont
  {Aab}} \emph {et~al.} (\bibinfo {collaboration} {Pierre Auger}),\ }\href
  {https://doi.org/10.1103/PhysRevD.90.122006} {\bibfield  {journal} {\bibinfo
  {journal} {Phys. Rev. D}\ }\textbf {\bibinfo {volume} {90}},\ \bibinfo
  {pages} {122006} (\bibinfo {year} {2014}{\natexlab{b}})},\ \Eprint
  {https://arxiv.org/abs/1409.5083} {arXiv:1409.5083 [astro-ph.HE]}
  \BibitemShut {NoStop}%
\bibitem [{\citenamefont {Kampert}\ and\ \citenamefont
  {Unger}(2012)}]{Kampert:2012mx}%
  \BibitemOpen
  \bibfield  {author} {\bibinfo {author} {\bibfnamefont {K.-H.}\ \bibnamefont
  {Kampert}}\ and\ \bibinfo {author} {\bibfnamefont {M.}~\bibnamefont
  {Unger}},\ }\href {https://doi.org/10.1016/j.astropartphys.2012.02.004}
  {\bibfield  {journal} {\bibinfo  {journal} {Astroparticle Physics}\ }\textbf
  {\bibinfo {volume} {35}},\ \bibinfo {pages} {660} (\bibinfo {year} {2012})},\
  \Eprint {https://arxiv.org/abs/1201.0018} {arxiv:1201.0018 [astro-ph.HE]}
  \BibitemShut {NoStop}%
\bibitem [{\citenamefont {Shibata}\ \emph {et~al.}(2010)\citenamefont
  {Shibata}, \citenamefont {Katayose}, \citenamefont {Huang},\ and\
  \citenamefont {Chen}}]{Shibata:2010zza}%
  \BibitemOpen
  \bibfield  {author} {\bibinfo {author} {\bibfnamefont {M.}~\bibnamefont
  {Shibata}}, \bibinfo {author} {\bibfnamefont {Y.}~\bibnamefont {Katayose}},
  \bibinfo {author} {\bibfnamefont {J.}~\bibnamefont {Huang}},\ and\ \bibinfo
  {author} {\bibfnamefont {D.}~\bibnamefont {Chen}},\ }\href
  {https://doi.org/10.1088/0004-637X/716/2/1076} {\bibfield  {journal}
  {\bibinfo  {journal} {Astrophys. J.}\ }\textbf {\bibinfo {volume} {716}},\
  \bibinfo {pages} {1076} (\bibinfo {year} {2010})}\BibitemShut {NoStop}%
\bibitem [{\citenamefont {Zhao}\ \emph {et~al.}(2015)\citenamefont {Zhao},
  \citenamefont {Jia},\ and\ \citenamefont {Zhu}}]{Zhao:2015rja}%
  \BibitemOpen
  \bibfield  {author} {\bibinfo {author} {\bibfnamefont {Y.}~\bibnamefont
  {Zhao}}, \bibinfo {author} {\bibfnamefont {H.-Y.}\ \bibnamefont {Jia}},\ and\
  \bibinfo {author} {\bibfnamefont {F.-R.}\ \bibnamefont {Zhu}},\ }\href
  {https://doi.org/10.1088/1674-1137/39/12/125001} {\bibfield  {journal}
  {\bibinfo  {journal} {Chin. Phys. C}\ }\textbf {\bibinfo {volume} {39}},\
  \bibinfo {pages} {125001} (\bibinfo {year} {2015})},\ \Eprint
  {https://arxiv.org/abs/1506.00352} {arXiv:1506.00352 [astro-ph.HE]}
  \BibitemShut {NoStop}%
\bibitem [{\citenamefont {Sveshnikova}\ \emph {et~al.}(2013)\citenamefont
  {Sveshnikova}, \citenamefont {Korosteleva}, \citenamefont {Kuzmichev},
  \citenamefont {Ptuskin}, \citenamefont {Prosin},\ and\ \citenamefont
  {Strelnikova}}]{Sveshnikova:2013qxa}%
  \BibitemOpen
  \bibfield  {author} {\bibinfo {author} {\bibfnamefont {L.~G.}\ \bibnamefont
  {Sveshnikova}}, \bibinfo {author} {\bibfnamefont {E.~E.}\ \bibnamefont
  {Korosteleva}}, \bibinfo {author} {\bibfnamefont {L.~A.}\ \bibnamefont
  {Kuzmichev}}, \bibinfo {author} {\bibfnamefont {V.~S.}\ \bibnamefont
  {Ptuskin}}, \bibinfo {author} {\bibfnamefont {V.~A.}\ \bibnamefont
  {Prosin}},\ and\ \bibinfo {author} {\bibfnamefont {O.~N.}\ \bibnamefont
  {Strelnikova}},\ }\href {https://doi.org/10.1088/1742-6596/409/1/012062}
  {\bibfield  {journal} {\bibinfo  {journal} {Journal of Physics: Conference
  Series}\ }\textbf {\bibinfo {volume} {409}},\ \bibinfo {pages} {012062}
  (\bibinfo {year} {2013})},\ \Eprint {https://arxiv.org/abs/1303.1713}
  {arxiv:1303.1713 [astro-ph.HE]} \BibitemShut {NoStop}%
\bibitem [{\citenamefont {Abu-Zayyad}\ \emph {et~al.}(2018)\citenamefont
  {Abu-Zayyad}, \citenamefont {Ivanov}, \citenamefont {Jui}, \citenamefont
  {Kim}, \citenamefont {Matthews}, \citenamefont {Smith}, \citenamefont
  {Thomas}, \citenamefont {Thomson},\ and\ \citenamefont
  {Zundel}}]{Abu-Zayyad:2018btv}%
  \BibitemOpen
  \bibfield  {author} {\bibinfo {author} {\bibfnamefont {T.}~\bibnamefont
  {Abu-Zayyad}}, \bibinfo {author} {\bibfnamefont {D.}~\bibnamefont {Ivanov}},
  \bibinfo {author} {\bibfnamefont {C.~C.~H.}\ \bibnamefont {Jui}}, \bibinfo
  {author} {\bibfnamefont {J.~H.}\ \bibnamefont {Kim}}, \bibinfo {author}
  {\bibfnamefont {J.~N.}\ \bibnamefont {Matthews}}, \bibinfo {author}
  {\bibfnamefont {J.~D.}\ \bibnamefont {Smith}}, \bibinfo {author}
  {\bibfnamefont {S.~B.}\ \bibnamefont {Thomas}}, \bibinfo {author}
  {\bibfnamefont {G.~B.}\ \bibnamefont {Thomson}},\ and\ \bibinfo {author}
  {\bibfnamefont {Z.}~\bibnamefont {Zundel}},\ }\href@noop {} {\bibfield
  {journal} {\bibinfo  {journal} {arXiv Eprint}\ } (\bibinfo {year} {2018})},\
  \Eprint {https://arxiv.org/abs/1803.07052} {arXiv:1803.07052 [astro-ph.HE]}
  \BibitemShut {NoStop}%
\bibitem [{\citenamefont {Deligny}(2020)}]{Deligny:2020gzq}%
  \BibitemOpen
  \bibfield  {author} {\bibinfo {author} {\bibfnamefont {O.}~\bibnamefont
  {Deligny}} (\bibinfo {collaboration} {Pierre Auger, Telescope Array}),\
  }\href {https://doi.org/10.22323/1.358.0234} {\bibfield  {journal} {\bibinfo
  {journal} {PoS}\ }\textbf {\bibinfo {volume} {ICRC2019}},\ \bibinfo {pages}
  {234} (\bibinfo {year} {2020})},\ \Eprint {https://arxiv.org/abs/2001.08811}
  {arXiv:2001.08811 [astro-ph.HE]} \BibitemShut {NoStop}%
\bibitem [{\citenamefont {Bergman}\ \emph {et~al.}(2024)\citenamefont {Bergman}
  \emph {et~al.}}]{PierreAuger:2023wti}%
  \BibitemOpen
  \bibfield  {author} {\bibinfo {author} {\bibfnamefont {D.~R.}\ \bibnamefont
  {Bergman}} \emph {et~al.} (\bibinfo {collaboration} {Pierre Auger, Telescope
  Array}),\ }\href {https://doi.org/10.22323/1.444.0406} {\bibfield  {journal}
  {\bibinfo  {journal} {PoS}\ }\textbf {\bibinfo {volume} {ICRC2023}},\
  \bibinfo {pages} {406} (\bibinfo {year} {2024})}\BibitemShut {NoStop}%
\bibitem [{\citenamefont {Plotko}\ \emph {et~al.}(2023)\citenamefont {Plotko},
  \citenamefont {van Vliet}, \citenamefont {Rodrigues},\ and\ \citenamefont
  {Winter}}]{Plotko:2022urd}%
  \BibitemOpen
  \bibfield  {author} {\bibinfo {author} {\bibfnamefont {P.}~\bibnamefont
  {Plotko}}, \bibinfo {author} {\bibfnamefont {A.}~\bibnamefont {van Vliet}},
  \bibinfo {author} {\bibfnamefont {X.}~\bibnamefont {Rodrigues}},\ and\
  \bibinfo {author} {\bibfnamefont {W.}~\bibnamefont {Winter}},\ }\href
  {https://doi.org/10.3847/1538-4357/acdf59} {\bibfield  {journal} {\bibinfo
  {journal} {Astrophys. J.}\ }\textbf {\bibinfo {volume} {953}},\ \bibinfo
  {pages} {129} (\bibinfo {year} {2023})},\ \Eprint
  {https://arxiv.org/abs/2208.12274} {arXiv:2208.12274 [astro-ph.HE]}
  \BibitemShut {NoStop}%
\bibitem [{\citenamefont {Abdul~Halim}\ \emph {et~al.}(2023)\citenamefont
  {Abdul~Halim} \emph {et~al.}}]{PierreAuger:2023yym}%
  \BibitemOpen
  \bibfield  {author} {\bibinfo {author} {\bibfnamefont {A.}~\bibnamefont
  {Abdul~Halim}} \emph {et~al.} (\bibinfo {collaboration} {Pierre Auger}),\
  }\href {https://doi.org/10.22323/1.444.0249} {\bibfield  {journal} {\bibinfo
  {journal} {PoS}\ }\textbf {\bibinfo {volume} {ICRC2023}},\ \bibinfo {pages}
  {249} (\bibinfo {year} {2023})}\BibitemShut {NoStop}%
\bibitem [{\citenamefont {Bergman}(2021)}]{Bergman:2021djm}%
  \BibitemOpen
  \bibfield  {author} {\bibinfo {author} {\bibfnamefont {D.}~\bibnamefont
  {Bergman}} (\bibinfo {collaboration} {Telescope Array}),\ }\href
  {https://doi.org/10.22323/1.395.0338} {\bibfield  {journal} {\bibinfo
  {journal} {PoS}\ }\textbf {\bibinfo {volume} {ICRC2021}},\ \bibinfo {pages}
  {338} (\bibinfo {year} {2021})}\BibitemShut {NoStop}%
\bibitem [{\citenamefont {Zhang}(2023)}]{Zhang:2023gbv}%
  \BibitemOpen
  \bibfield  {author} {\bibinfo {author} {\bibfnamefont {S.}~\bibnamefont
  {Zhang}} (\bibinfo {collaboration} {LHAASO}),\ }\href
  {https://doi.org/10.22323/1.444.0408} {\bibfield  {journal} {\bibinfo
  {journal} {PoS}\ }\textbf {\bibinfo {volume} {ICRC2023}},\ \bibinfo {pages}
  {408} (\bibinfo {year} {2023})}\BibitemShut {NoStop}%
\bibitem [{\citenamefont {Cattaneo}(2019)}]{Cattaneo:2019uui}%
  \BibitemOpen
  \bibfield  {author} {\bibinfo {author} {\bibfnamefont {P.~W.}\ \bibnamefont
  {Cattaneo}} (\bibinfo {collaboration} {HERD}),\ }\href
  {https://doi.org/10.1016/j.nuclphysbps.2019.07.013} {\bibfield  {journal}
  {\bibinfo  {journal} {Nucl. Part. Phys. Proc.}\ }\textbf {\bibinfo {volume}
  {306-308}},\ \bibinfo {pages} {85} (\bibinfo {year} {2019})}\BibitemShut
  {NoStop}%
\bibitem [{\citenamefont {Olinto}\ \emph {et~al.}(2021)\citenamefont {Olinto}
  \emph {et~al.}}]{POEMMA:2020ykm}%
  \BibitemOpen
  \bibfield  {author} {\bibinfo {author} {\bibfnamefont {A.~V.}\ \bibnamefont
  {Olinto}} \emph {et~al.} (\bibinfo {collaboration} {POEMMA}),\ }\href
  {https://doi.org/10.1088/1475-7516/2021/06/007} {\bibfield  {journal}
  {\bibinfo  {journal} {JCAP}\ }\textbf {\bibinfo {volume} {06}},\ \bibinfo
  {pages} {007}},\ \Eprint {https://arxiv.org/abs/2012.07945} {arXiv:2012.07945
  [astro-ph.IM]} \BibitemShut {NoStop}%
\bibitem [{\citenamefont {Di~Santo}(2023)}]{DiSanto:2023lcx}%
  \BibitemOpen
  \bibfield  {author} {\bibinfo {author} {\bibfnamefont {M.}~\bibnamefont
  {Di~Santo}},\ }\href {https://doi.org/10.1142/S2010194523610049} {\bibfield
  {journal} {\bibinfo  {journal} {Int. J. Mod. Phys. Conf. Ser.}\ }\textbf
  {\bibinfo {volume} {51}},\ \bibinfo {pages} {2361004} (\bibinfo {year}
  {2023})}\BibitemShut {NoStop}%
\bibitem [{\citenamefont {Aartsen}\ \emph {et~al.}(2021)\citenamefont {Aartsen}
  \emph {et~al.}}]{IceCube-Gen2:2020qha}%
  \BibitemOpen
  \bibfield  {author} {\bibinfo {author} {\bibfnamefont {M.~G.}\ \bibnamefont
  {Aartsen}} \emph {et~al.} (\bibinfo {collaboration} {IceCube-Gen2}),\ }\href
  {https://doi.org/10.1088/1361-6471/abbd48} {\bibfield  {journal} {\bibinfo
  {journal} {J. Phys. G}\ }\textbf {\bibinfo {volume} {48}},\ \bibinfo {pages}
  {060501} (\bibinfo {year} {2021})},\ \Eprint
  {https://arxiv.org/abs/2008.04323} {arXiv:2008.04323 [astro-ph.HE]}
  \BibitemShut {NoStop}%
\bibitem [{\citenamefont {Fujii}\ \emph {et~al.}(2016)\citenamefont {Fujii}
  \emph {et~al.}}]{Fujii:2015dra}%
  \BibitemOpen
  \bibfield  {author} {\bibinfo {author} {\bibfnamefont {T.}~\bibnamefont
  {Fujii}} \emph {et~al.},\ }\href
  {https://doi.org/10.1016/j.astropartphys.2015.10.006} {\bibfield  {journal}
  {\bibinfo  {journal} {Astropart. Phys.}\ }\textbf {\bibinfo {volume} {74}},\
  \bibinfo {pages} {64} (\bibinfo {year} {2016})},\ \Eprint
  {https://arxiv.org/abs/1504.00692} {arXiv:1504.00692 [astro-ph.IM]}
  \BibitemShut {NoStop}%
\bibitem [{\citenamefont {de~Mello~Neto}(2023)}]{deMelloNeto:2023zvk}%
  \BibitemOpen
  \bibfield  {author} {\bibinfo {author} {\bibfnamefont {J.~a. R.~T.}\
  \bibnamefont {de~Mello~Neto}} (\bibinfo {collaboration} {GRAND}),\ }\href
  {https://doi.org/10.22323/1.444.1050} {\bibfield  {journal} {\bibinfo
  {journal} {PoS}\ }\textbf {\bibinfo {volume} {ICRC2023}},\ \bibinfo {pages}
  {1050} (\bibinfo {year} {2023})},\ \Eprint {https://arxiv.org/abs/2307.13638}
  {arXiv:2307.13638 [astro-ph.HE]} \BibitemShut {NoStop}%
\bibitem [{\citenamefont {Alves~Batista}(2023)}]{AlvesBatista:2023lqg}%
  \BibitemOpen
  \bibfield  {author} {\bibinfo {author} {\bibfnamefont {R.}~\bibnamefont
  {Alves~Batista}} (\bibinfo {collaboration} {GCOS}),\ }\href
  {https://doi.org/10.22323/1.444.0281} {\bibfield  {journal} {\bibinfo
  {journal} {PoS}\ }\textbf {\bibinfo {volume} {ICRC2023}},\ \bibinfo {pages}
  {281} (\bibinfo {year} {2023})},\ \Eprint {https://arxiv.org/abs/2309.17324}
  {arXiv:2309.17324 [astro-ph.HE]} \BibitemShut {NoStop}%
\end{thebibliography}%

%
\end{document}